\documentclass[12pt]{article}

\newcommand{\rem}[1]{}
\newcommand{\remv}[1]{}

\usepackage{amsmath}

\usepackage{graphicx}

\newcommand{\leftsup}[2]{{\vphantom{#2}}^{#1\!}{#2}} 
\newcommand{\leftsub}[2]{{\vphantom{#2}}_{#1\!}{#2}}
\def\upar{\mbox{\tiny  \bf up}}
\def\dnar{\mbox{\tiny  \bf dn}}
\newcommand{\urmbf}[1]{ \! {\rm \bf  #1} }

\begin{document}

\begin{titlepage}
\begin{flushright}

\end{flushright}

\begin{center}
{\Large\bf $ $ \\ $ $ \\
Finite dimensional vertex 
}\\
\bigskip\bigskip\bigskip
{\large Andrei Mikhailov\footnote{On leave from 
Institute for Theoretical and 
Experimental Physics, 
117259, Bol. Cheremushkinskaya, 25, 
Moscow, Russia}}
\\
\bigskip\bigskip
{\it Instituto de F\'{i}sica Te\'orica, Universidade Estadual Paulista\\
R. Dr. Bento Teobaldo Ferraz 271, 
Bloco II -- Barra Funda\\
CEP:01140-070 -- S\~{a}o Paulo, Brasil\\
}

\vskip 1cm
\end{center}

\begin{abstract}
The spectrum of linearized excitations of the Type IIB SUGRA on $AdS_5\times S^5$
contains both unitary and non-unitary representations. Among the non-unitary, 
some are finite-dimensional. We explicitly construct the pure spinor  vertex 
operators for a family of such finite-dimensional representations. 
The construction can also be applied to infinite-dimensional representations, 
including unitary, although it becomes in this case somewhat less explicit. 
\end{abstract}

\end{titlepage}

\tableofcontents

\section{Introduction}
The maximally supersymmetric 10-dimensional background $AdS_5\times S^5$  \cite{Romans:1984an} of 
the 
Type IIB superstring is of the crucial importance in modern string theory, 
especially in the context of AdS/CFT correspondence. For various reasons, it 
is important to understand infinitesimal deformations of this background. 
They 
correspond to small fluctuations of the classical supergravity fields around 
their ``vacuum'' values in $AdS_5\times S^5$. First of all, one can study them in 
the linearized approximation, to the first order in the small parameter 
describing the deviation of the solution from $AdS_5\times S^5$. We can call such 
solutions ``linearized excitations of $AdS_5\times S^5$''. These linearized 
excitations can be normalizable and non-normalizable. The normalizable ones 
correspond to  states of the ${\cal N} =4$ SYM on ${\bf R}\times S^3$, and the non-normalizable 
to deformations of the ${\cal N} = 4$ SYM \cite{Witten:1998qj}. Notice that the symmetry group of 
$AdS_5\times S^5$ naturally acts on the space of linearized excitations. It is 
natural to ask the following question:
\begin{itemize}
\item in which representations of the symmetry group of $AdS_5\times S^5$
   do these linearized excitations transform?
\end{itemize}
For the normalizable excitations, the answer is well-known; it is equivalent
to the classification of the local half-BPS operators in ${\cal N}=4$ SYM. The 
research program to classify them was initiated in \cite{Kim:1985ez}.   The answer is a 
series of unitary representations parametrized by a positive integer. 
Mathematically, the ${\cal N}=4$ superspace is a super-Grassmannian of embeddings:
\begin{equation}
{\bf C}^{2|2}\subset {\bf C}^{2+2|4}
\end{equation}
and the half-BPS operators are holomorphic sections of the $n$-th power of the 
Berezinian line bundle. See \cite{Howe:2001je} and references therein.

For the non-normalizable excitations, the situation is more complicated.
The space of non-normalizable excitations  is not an irreducible 
representation of the superconformal group, not even of the conformal group. 
There are subspaces which do not have an invariant complement. To the best 
of our knowledge, the representation content of the non-normalizable 
excitations has not been worked out.

By AdS/CFT the non-normalizable excitations correspond to the deformations 
of the SYM action  \cite{Witten:1998qj}:
\begin{equation}
S_{YM} \to S_{YM} + \varepsilon \int d^4x \;\rho(x)\; {\cal O}_{\Delta}(x)
\end{equation}
where ${\cal O}$ is a local operator of conformal dimension $\Delta$  and $\rho(x)$ 
is a density of the conformal weight $\Delta - 4$. Interestingly,  for any integer 
$\Delta \geq 4$ the space of densities has a {\em finite-dimensional subspace} 
invariant under the conformal group $SO(2,4)$. Acting on this space by the
supersymmetries we generate a finite-dimensional representation of the
full superconformal group $PSU(2,2|4)$.

In this paper we will construct the pure spinor vertex operators 
corresponding to some of these finite-dimensional spaces. There are two main
motivations. First of all, given the importance of the AdS background in 
string theory, we would like to know the complete spectrum of the linearized 
SUGRA on this background, not just unitary representations. Second, from the 
point of view of the pure spinor formalism, the classification of the 
finite-dimensional vertices is equivalent to a problem in linear algebra, 
which is interesting in itself. This may  be also related to the 
${\cal N} = 4$ integrability program along the lines of \cite{Mikhailov:2009rx}.

There were two previously known examples of a finite-dimensional vertex: the 
zero mode of the dilaton constructed in \cite{Berkovits:2008ga}, and the vertex for the 
beta-deformation considered in \cite{Bedoya:2010qz}. Here we construct an infinite series of 
new examples.

\paragraph     {Plan of the paper}
We will construct the universal vertex in the sense of \cite{Mikhailov:2009rx} for a specific
finite-dimensional representation. We describe this representation in 
Section \ref{sec:AlgebraicPreliminaries}. Then in Section \ref{sec:FieldTheoryPointOfView} we discuss (as a conjecture) 
the SYM interpretation. The construction itself is described in
Secions \ref{sec:BRSTAndParabolicInduction}, \ref{sec:TwoZeroPart}, \ref{sec:OneOnePart} and \ref{sec:GaugeChoices}. A partial analysis of the corresponding
SUGRA solutions is presented in Section \ref{sec:BFieldAndDilaton}. A possible generalization to 
infinite-dimensional (including unitary) representation is discussed in 
Section \ref{sec:Generalization}. In Section \ref{sec:SubspacesAndFactorspaces} we  discuss some representation-theoretic 
properties of our construction. Open questions are listed in Section \ref{sec:OpenQuestions}.

\section{Algebraic preliminaries}\label{sec:AlgebraicPreliminaries}
\subsection{Ansatz for the vertex}
We will use the following ansatz for the vertex transforming in the
representation ${\cal H}$ of ${\bf g}$. For every $\Psi\in {\cal H}$ the corresponding vertex is \cite{Mikhailov:2009rx}:
\begin{equation}\label{Ansatz}
V[\Psi](g,\lambda) = \left\langle v(\lambda) \;,\; g \Psi \right\rangle
\end{equation}
where $v(\lambda)\in {\cal H}'$ --- a constant ({\it i.e.} independent on $g$) vector 
in ${\cal H}'$.  Here we denote ${\cal H}'$ the space dual to ${\cal H}$:
\begin{equation}
{\cal H}' = \mbox{Hom}_{\bf C}({\cal H},{\bf C})
\end{equation}

\subsection{Definition of ${\cal H}$}
We will construct $v(\lambda)$ for a specific series of finite-dimensinonal 
representations $\cal H$, which we will now define.

\subsubsection{Parabolic induction}
\paragraph     {Block structure of $sl(4|4)$}
The even subalgebra of $sl(4|4)$ is a direct sum of two Lie algebras:
\begin{equation}
{\bf g}_{even} = {\bf g}_{\upar} \oplus {\bf g}_{\dnar}
\end{equation}
(The real form would be ${\bf g}_{\upar} = {\bf u(2,2)}$ and ${\bf g}_{\dnar} = {\bf u(4)}$.) Schematically,
in the $4\times 4$-block notations:
\begin{equation}
{\bf g} = \left[ \begin{array}{cc} 
      {\bf g}_{\upar}  & {\bf n}_+ \cr
      {\bf n}_-         & {\bf g}_{\dnar}
\end{array}\right]
\end{equation}
The $\bf n_-$ in the upper right corner and ${\bf n}_+$ in the lower left corner are both
odd abelian subalgebras ${\bf C}^{0|16}$.

\paragraph     {Parabolic subalgebra $\bf p$}
Let us denote ${\bf p}$  the following parabolic subalgebra of ${\bf g}$:
\begin{equation}\label{ParabolicSubalgebra}
{\bf p} = \left[ \begin{array}{cc} 
      {\bf g}_{\upar}    &  0 \cr
      {\bf n}_-         & {\bf g}_{\dnar}
\end{array}\right]
\end{equation}
We will denote:
\begin{align}
F_{\upar} \;:\;& \mbox{the fundamental of } {\bf g}_{\upar}
\\   
F_{\dnar} \;:\;& \mbox{the fundamental of } {\bf g}_{\dnar}
\end{align}
and $F'_{\upar}$, $F'_{\dnar}$ will denote the corresponding dual representations (a.k.a. 
``antifundamental representations''). 

\subsubsection{Construction of ${\cal H}$ as an induced representation}\label{sec:HAsInduced}
We will construct a series of representations of ${\bf g}$ from a series
of finite-dimensional representations of the bosonic subalgebra ${\bf g}_{\upar} \oplus {\bf g}_{\dnar}$,
using the parabolic induction. 

\paragraph     {A series of representations of the bosonic algebra ${\bf g}_{\rm even}$}
Let us consider the following finite 
dimensional representation of ${\bf g}_{\upar} \oplus {\bf g}_{\dnar}$ parametrized by an integer $n$:
\begin{equation}
L = Y\left( F_{\dnar}^{\otimes 2(n+1)}\right) 
\otimes Y\left((F'_{\upar})^{\otimes 2(n+1)} \right)
\end{equation}
where $Y$ is some specific Young symmetrizer, which acts  as follows.
The space $F_{dn}^{\otimes 2(n+1)}$ consists of the  tensors $f^{a_1\cdots a_{2(n+1)}}$;
the operation $Y$ first antisymmetrizes $[a_1a_2],\ldots,[a_{2n+1}a_{2n+2}]$, 
and then symmetrizes $(a_1a_3a_5\cdots a_{2n+1})$ and $(a_2a_4a_6\cdots a_{2n+2})$.
Similarly the space $Y(F'_{\upar})^{\otimes 2(n+1)}$ is the space of tensors  $f_{\alpha_1\cdots \alpha_{2(n+1)}}$ 
which are antisymmetrized and then symmetrized in the same way.

\paragraph     {Induced representation of ${\bf g}$}
Let us extend $L$ to a representation of the parabolic subalgebra (\ref{ParabolicSubalgebra}) by 
letting ${\bf n}_-$ act as $0$.

We will define: 
\begin{equation}\label{DefH}
{\cal H} = \mbox{Ind}_{\bf p}^{\bf g} \; L
\end{equation}
where the operation $\mbox{Ind}_{\bf p}^{\bf g}$ is defined as follows:
\begin{equation}\label{VertexRepresentationInduced}
\mbox{Ind}_{\bf p}^{\bf g} \; L = {\cal U}{\bf g} \otimes_{\bf p} L
\end{equation}
The nilpotent subalgebra ${\bf n}_-$ acts trivially on $L$.

\paragraph     {Tensor notations}
The elements of $L$ are tensors:
\begin{equation}
V{}^{a_1\ldots a_{2n+2}}_{\alpha_1\ldots \alpha_{2n+2}}
\end{equation}
with the appropriate symmetry conditions. 
We will introduce the coordinates $(\theta_+)^{\alpha}_a$ on ${\bf n}_+$.

\subsection{Dual space ${\cal H}'$ and holomorphic vector bundles on $G/G_{\rm even}$}
\label{sec:HolomorphicVectorBundles}
\subsubsection{Duality between induced and coinduced representations}
Notice that $v(\lambda)$ in Eq. (\ref{Ansatz}) belongs to the representation dual to ${\cal H}$. 
The dual representation to $\mbox{Ind}_{\bf p}^{\bf g}L$ is the coinduced representation:
\begin{equation}
{\cal H}' =\mbox{Coind}_{\bf p}^{\bf g}\; \;L' =
\mbox{Hom}_{\bf p}\left({\cal U}{\bf g} \; , \; L'\right)
\end{equation}
We consider both ${\cal U}{\bf g}$ and $L'$ as right ${\bf p}$-modules;
${\cal H}'$ can be thought as the space of functions $f:\;{\cal U}{\bf g} \to L'$ satisfying the 
property: 
\begin{equation}
f(\xi\eta) = f(\xi)\eta \;\;
\mbox{\small for $\xi\in {\cal U}{\bf g}$ and $\eta \in {\cal U}{\bf p}$ }
\end{equation}
The duality pairing is:
\begin{equation}
\langle f ,\; \xi\otimes_{{\cal U}{\bf p}} l \rangle = 
\langle f(\xi),l \rangle
\end{equation}
where $\langle,\rangle$ on the RHS is the pairing between $L'$ and $L$.
\remv{Details about pairing:}\rem{ photo/shot0009.png }

\paragraph     {Consistency of the definition of pairing}
Take some $\xi\in {\cal U}{\bf g}$, $g_{\rm even}\in {\cal U}{\bf g}_{\rm even}$ and $g_-\in {\cal U}{\bf n}_-$. We get:
\begin{equation}
\left\langle \Phi_{\lambda}(\xi g_{\rm even} g_- )\; , \; l \right\rangle =
\left\langle \Phi_{\lambda}(\xi) g_{\rm even} g_- \; , \; l\right\rangle =
\left\langle \Phi_{\lambda}(\xi) \; , \; g_{\rm even}l \right\rangle
\end{equation}
\paragraph     {Invariance under global rotations}
\begin{equation}\label{GlobalRotations}
\left\langle g.\Phi_{\lambda} \; , \; \xi\otimes_{\bf p} l \right\rangle =
\left\langle \Phi_{\lambda}(g^{-1}\xi) \; , \; l \right\rangle =
\left\langle \Phi_{\lambda} \; , \; g^{-1}\xi\otimes_{\bf p} l \right\rangle
\end{equation}
\paragraph     {${\bf g}_0$-covariance}
In particular, consider the case when $g\in {\cal U}{\bf g}_{\bar{0}}$ in (\ref{GlobalRotations}):
\begin{equation}
(h.\Phi_{\lambda})(\xi) = \Phi_{\lambda}(h^{-1}\xi) = 
\Phi_{\lambda}(h^{-1}\xi h)h^{-1} 
\end{equation}
For covariance, we want this to be equal to $\Phi_{h\lambda h^{-1}}(\xi)$. Therefore we need
to impose the ${\bf g}_{\bar{0}}$-{\em covariance condition} on $\Phi$:
\begin{equation}
\Phi_{\lambda}(\xi) = \Phi_{h^{-1}\lambda h} (h^{-1}\xi h) h^{-1}
\end{equation}

\subsubsection{Relation to \cite{Berkovits:2007rj}}
Geometrically ${\cal H}'$ can be thought of as the space of holomorphic sections of 
the vector bundle on the odd Grassmanian $G/G_{\rm even}$ with the fiber $L'$. Notice 
that $G/G_{\rm even}$ is the target space of the  gauged linear sigma-model of \cite{Berkovits:2007rj}.
Although that was our main motivation in using the parabolic induction, the 
precise relation of our method to the discussion in \cite{Berkovits:2007rj} is not clear to us. 

\subsubsection{Explicit formulas for the action of global symmetries}
Consider $\eta_+\in {\bf n}_+$ and $\eta_-\in {\bf n}_-$. We get:
\begin{align}
e^{-\eta_+}\Phi\left(e^{\theta_+}\right) = & \; 
\Phi\left(e^{\eta_+ + \theta_+}\right)
\label{ShiftOfThetaPlus} \\     
\left.{d\over dt}\right|_{t=0} e^{-t\eta_-}\Phi\left(e^{\theta_+}\right) = & \;
\left.{d\over dt}\right|_{t=0} \Phi\left(e^{t\eta_-} e^{\theta_+}\right) = 
\label{ActionOfQMinus}
\\    
= &\; 
 {1\over 2} \left(
   [\theta_+,[\theta_+,\eta_-]] {\partial\over\partial\theta_+}
\right) 
\Phi\left( e^{\theta_+}\right)
+ \; \Phi\left( e^{\theta_+}\right) \;
[\eta_-,\theta_+] 
\nonumber
\end{align}

\subsection{Coinduced representation is not irreducible}\label{sec:RelatedT}
Let us denote $F$ (without lower index) the fundamental representation of 
the {\em super}-algebra ${\bf g}$, and $F'$ the dual (antifundamental) representation.
Let us consider the following symmetrized tensor product:
\begin{equation}
\label{DefT}
{\cal T} = Y(F^{\otimes 2(n+1)})
\otimes Y^{\rm tr}((F')^{\otimes 2(n+1)})
\end{equation}
where $Y$ and $Y^{\rm tr}$ are the super-symmetrizers. There is a canonical map:
\begin{equation}
\mbox{ev }: \;  {\cal H} \to {\cal T}
\end{equation}
It is constructed using the  embedding $\iota$: 
\begin{equation}\label{IotaEmbedding}
 Y\left( (F'_{\upar})^{\otimes 2(n+1)}\right) 
\otimes Y\left((F_{\dnar})^{\otimes 2(n+1)} \right) \;\; \stackrel{\iota}{\longrightarrow} \;\;
Y(F^{\otimes 2(n+1)})\otimes Y^{\rm tr}((F')^{\otimes 2(n+1)})
\end{equation}
Using this embedding, $\mbox{ev}$ is defined as the action of the element of
${\cal U}{\bf g}$ on the embedded tensor:
\begin{equation}
\mbox{ev }(\xi\otimes f) = \xi \iota(f)
\end{equation}
Notice that $\mbox{ker}(\mbox{ev})\subset {\cal H}$ is an invariant subspace, but there is no 
complementary invariant subspace. Therefore, ${\cal H}'$ has an invariant subspace 
consisting of those functionals which vanish on $\mbox{ker}(\mbox{ev})$. This subspace 
consists of the following functionals, using the notations of Section \ref{sec:HolomorphicVectorBundles}:
\begin{equation}\label{ImageOfMultiplication}
f(\xi) =  \left.\left(T \xi \right)\right|_{{\rm restriction\; to\; }L}
\end{equation}
where $T\in  {\cal T}'$ --- a linear function on $\cal T$. On the right hand side we 
evaluate the action of $\xi\in {\cal U}{\bf g}$ on this $T$, and then restrict the resulting 
linear functional to $L\subset {\cal T}$, where the embedding of $L$ into $\cal T$ is the $\iota$  of 
(\ref{IotaEmbedding}). We will denote the subspace of functions of the form (\ref{ImageOfMultiplication}) in the 
standard way:
\begin{equation}\label{DefKerEvPerp}
\left(\mbox{ker}(\mbox{ev})\right)^{\perp} = 
\left\{ 
f\; |\; f(\xi) =  \left.\left(T\xi\right)\right|_{{\rm restriction\; to\; }L}
\right\}
\end{equation}
More explicitly, given a tensor $T^{i_1i_2\cdots i_{2n}}_{j_1j_2\cdots j_{2n}}$, we associate to it the following 
holomorphic section:
\begin{equation}\label{HolomorphicInduction}
\Phi[T]^{\alpha_1\cdots \alpha_{2n}}_{a_1\cdots a_{2n}} (\theta_+) = 
\left( Te^{\theta_+} \right)^{\alpha_1\cdots \alpha_{2n}}_{a_1\cdots a_{2n}} 
\end{equation}
Such sections form an invariant subspace  $(\mbox{ker(ev)})^{\perp}\subset \mbox{Coind}_{\bf p}^{\bf g}L'$.

\subsection{Properties of $Y\left(F^{\otimes 2(n+1)}\right)$}
\paragraph     {Dimension}
The representation $Y\left(F^{\otimes 2(n+1)}\right)$ can be identified with the 
space of traceless symmetric tensors of $so(6)$. The dimension is:
\remv{Calculation of hooks:}\rem{ photo/shot0020.png }
\remv{Dimensions of symmetric traceless tensors:}\rem{ photo/shot0021.png }
\begin{equation}
\mbox{dim}\;Y\left(F^{\otimes 2(n+1)}\right) = {(n+2)(n+3)^2(n+4)\over 12}
\end{equation}
For example, when $n=1$ we get:
\begin{equation}
{\begin{picture}(11,21)(0,4)
\multiput(0,0)(0,10){2}{\framebox(10,10){}}
\end{picture}
\otimes 
\begin{picture}(11,21)(0,4)
\multiput(0,0)(0,10){2}{\framebox(10,10){}}
\end{picture} 
\atop 36}
\;\; = \;\;
{\begin{picture}(11,41)(0,4)
\multiput(0,0)(0,10){4}{\framebox(10,10){}}
\end{picture} 
\atop 1}
\;  +  \; 
{\begin{picture}(21,31)(0,4)
\multiput(0,0)(0,10){3}{\framebox(10,10){}}
\put(10,20){\framebox(10,10){}}
\end{picture} 
\atop 15}
\;  +  \; 
{\begin{picture}(21,21)(0,4)
\multiput(0,0)(0,10){2}{\framebox(10,10){}}
\multiput(10,0)(0,10){2}{\framebox(10,10){}}
\end{picture} 
\atop 20}
\end{equation}
The symmetrization of $(\alpha_1\alpha_3)$ and $(\alpha_2\alpha_4)$ removes ${\bf 15}$ and  ${\bf 1}$ and leaves ${\bf 20}$.

\paragraph     {Decomposition as a representation of $sp(2)=so(5)$}
The tensor product of $n+1$ vector representations of $so(5)$ has
an invariant subspace $T_{n+1}$, which consists of the symmetric traceless
tensors. Its dimension is $
{(2n+5)(n+2)(n+3)\over 6}
$. We have the decomposition:
\begin{align}
Y\left(F^{\otimes 2(n+1)}\right) = &\; \bigoplus\limits_{k=0}^{n+1} T_k
\\     
{(n+2)(n+3)^2(n+4)\over 12} = &\; \sum\limits_{k=0}^{n+1}
{(2n+5)(n+2)(n+3)\over 6}
\end{align}

\section{Field theory point of view}\label{sec:FieldTheoryPointOfView}
\subsection{A finite-dimensional supermultiplet of deformations}
We conjecture that our vertex operator is dual to the finite-dimensional
supermultiplet of  deformations of the ${\cal N}=4$ SYM generated by the following
deformation:
\begin{equation}
S_{YM} \to S_{YM} + \varepsilon \int d^4x \;\rho(x)\; \mbox{tr}\;Z^{n+3}(x)
\end{equation}
where $Z = \Phi^5 + i \Phi^6$ is a complex combination of the SYM scalars and $\rho(x)$ 
is a density of the conformal weight $n-1$. Observe that for any integer 
$n>0$ the space of densities has a finite-dimensional subspace invariant
under the conformal group $SO(2,4)$. This can be seen for example from 
the dual AdS picture. In the AdS language, this subspace corresponds to the
harmonic polynomials of weight $n-1$ in ${\bf R}^{2+4}$. This can be also seen
as follows.  The infinitesimal special conformal transformations act on $Z$
as follows:
\begin{equation}
\delta_K Z(x) = \left( 
   (k\cdot x) (x\cdot \partial) - {1\over 2}(x\cdot x) (k\cdot\partial)
   + (k\cdot x)
\right) Z(x)
\end{equation}
Therefore the deformation is invariant with the following transformation
rule\footnote{We could think of $\rho(x)$ as a ``spectator field'', or 
as an $x$-dependent coupling constant} for $\rho(x)$:
\begin{equation}
\delta_K \rho(x) = \left( 
   (k\cdot x) (x\cdot \partial) - {1\over 2}(x\cdot x) (k\cdot\partial)
   + (1-n)(k\cdot x)
\right) \rho(x)
\end{equation}
In other words, we should think of $\rho$ as a ``density'' of weight $1-n$.
Consider the linear space consisting of the densities $\rho(x)$ of the following
form:
\begin{equation}\label{FiniteDimensional}
\rho(x) = \sum_{m=0}^{n-1} q_{n-m-1}\left((x\cdot x)\right) p_m(x)
\end{equation}
where $p_m(x)$ is a homogeneous polynomial in $x$ of weight $m$, and
$q_{n-m}\left((x\cdot x)\right)$ a polynomial in $(x\cdot x)$ of weight $n-m$ (not necessarily 
homogeneous; {\it e.g.} $q\left((x\cdot x)\right) = 1$ is OK). One can see this space is
closed under the action of the conformal transformations. This means that
the (infinite-dimensional) representation of the conformal group $su(2,2)$ 
in the space of densities of the weight $1-n$ for $n>0$ has an invariant
finite-dimensional subspace\footnote{There is no invariant 
complementary subspace, therefore the space of densities is not a
semisimple representation of $su(2,2)$.} (\ref{FiniteDimensional}). 

For example, when $n=1$ there is a conformally invariant deformation:
\begin{equation}\label{DeformationCaseZ4}
S_{YM} \to S_{YM} + \varepsilon \int \; d^4 x \;\mbox{tr}\; Z^4
\end{equation}
Let us consider a particular case $n=1$, which corresponds to $\mbox{tr}\; Z^4$.

\subsection{Action of R-symmetry on $\mbox{tr}\;Z^k$}
Let us first consider the action of $so(6)$ on (\ref{DeformationCaseZ4}). 
The perturbation transforms in the following representation of $so(6)$:
\begin{equation}\label{DiagrammZ4}
\begin{picture}(40,20)(0,2)
\multiput(0,0)(10,0){4}{
  \multiput(0,0)(0,10){2}{\framebox(10,10){}}
}
\end{picture}
\end{equation}
which has dimension {\bf 105}. The highest weight is:
\begin{equation}\label{DiagrammZ4Labels}
\begin{picture}(40,20)(0,2)
\multiput(0,0)(10,0){4}{
  \put(0,10){\framebox(10,10){\tiny 1}}
  \put(0,0){\framebox(10,10){\tiny 2}}
} 
\end{picture}
\end{equation}
Similarly:
\begin{equation}
\mbox{tr}\; Z^5 :\; 
\begin{picture}(50,20)(0,2)
\multiput(0,0)(10,0){5}{
  \put(0,10){\framebox(10,10){\tiny 1}}
  \put(0,0){\framebox(10,10){\tiny 2}}
} 
\end{picture}
\qquad
\mbox{tr}\; Z^6:\;
\begin{picture}(60,20)(0,2)
\multiput(0,0)(10,0){6}{
  \put(0,10){\framebox(10,10){\tiny 1}}
  \put(0,0){\framebox(10,10){\tiny 2}}
} 
\end{picture}
\qquad \mbox{etc.}
\end{equation}

For $\mbox{tr}\; Z^4$ there are states constant in AdS. They correspond to:
\begin{equation}
\left(E_{a_1}^{\alpha'}E_{a_2}^{\beta'}E_{a_3}^{\gamma'}E_{a_4}^{\delta'} \right)
\mbox{\large $\otimes$} \left(
Y\left(e_{\alpha'} \otimes e_{\beta'}\otimes e_{\gamma'}\otimes e_{\delta'}\right)
\otimes
\widetilde{Y}\left(e^{b_1} \otimes e^{b_2}\otimes e^{b_3}\otimes e^{b_4}\right)
\right)
\end{equation}
The state corresponding to $\int \; d^4 x \;\mbox{tr}\; Z^4$ is obtained by taking
$(a_1,a_2,a_3,a_4) = (3,4,3,4)$ and $(b_1,b_2,b_3,b_4) = (1,2,1,2)$.

\subsection{Supersymmetric Young diagramms}
The diagramm (\ref{DiagrammZ4Labels}) can be embedded in the following supersymmetric Young 
diagramm\footnote{Notice that the SUSY Young diagramms have been considered
previously in the context of AdS/CFT in \cite{Dasgupta:2002ru}, but in 
that paper only the products of fundamental representations are needed. Here
we discuss Young diagramms involving both fundamental and antifundamental
representations.}, using the notations of \cite{MoensSchurFunctions}:
\begin{equation}
\begin{picture}(40,40)(0,2)
\multiput(0,20)(10,0){2}{
  \put(0,10){\framebox(10,10){\tiny 3}}
  \put(0,0){\framebox(10,10){\tiny 4}}
} 
\multiput(20,0)(10,0){2}{
  \put(0,10){\framebox(10,10){\tiny 1}}
  \put(0,0){\framebox(10,10){\tiny 2}}
} 
\end{picture}
\end{equation}
(The labels correspond to the highest weight state.) Similarly, we have
the following diagramms for $\mbox{tr}\;Z^5$, $\mbox{tr}\;Z^6$ {\it etc.}:
\begin{equation}
\begin{picture}(50,50)(0,2)
\put(0,20){
  \put(0,20){\framebox(10,10){\tiny $1_{\rm F}$}}
  \put(0,10){\framebox(10,10){\tiny 3}}
  \put(0,0){\framebox(10,10){\tiny 4}}
} 
\put(10,20){
  \put(0,20){\framebox(10,10){\tiny $2_{\rm F}$}}
  \put(0,10){\framebox(10,10){\tiny 3}}
  \put(0,0){\framebox(10,10){\tiny 4}}
} 
\multiput(20,0)(10,0){3}{
  \put(0,10){\framebox(10,10){\tiny 1}}
  \put(0,0){\framebox(10,10){\tiny 2}}
} 
\end{picture}
\hspace{40pt}
\begin{picture}(60,60)(0,2)
\put(0,20){
  \put(0,30){\framebox(10,10){\tiny $1_{\rm F}$}}
  \put(0,20){\framebox(10,10){\tiny $1_{\rm F}$}}
  \put(0,10){\framebox(10,10){\tiny 3}}
  \put(0,0){\framebox(10,10){\tiny 4}}
} 
\put(10,20){
  \put(0,30){\framebox(10,10){\tiny $2_{\rm F}$}}
  \put(0,20){\framebox(10,10){\tiny $2_{\rm F}$}}
  \put(0,10){\framebox(10,10){\tiny 3}}
  \put(0,0){\framebox(10,10){\tiny 4}}
} 
\multiput(20,0)(10,0){4}{
  \put(0,10){\framebox(10,10){\tiny 1}}
  \put(0,0){\framebox(10,10){\tiny 2}}
} 
\end{picture}
\end{equation}
where $1_{\rm F}$ means that the index is fermionic, and therefore the highest
weight state for $\mbox{tr}\;Z^n$ has to have a nontrivial dependence on spacetime 
coordinates, for $n>4$. 

However, our vertex corresponds to the indecomposable representation
(\ref{VertexRepresentationInduced}) (the Kac module) rather than the irreducible representation 
corresponding to the Yound diagramm.

\subsection{A puzzle}
Now let us consider $\mbox{tr}\; Z^{k}$ with $k>4$. The lowest possible AdS momentum 
is then $k-4$. However, it appears that we can construct vertices with
lower momentum. For example, consider the following state in the multiplet
which should correspond to  $\mbox{tr}\;Z^5$:
\begin{align}
& \left(
   E_{a_1}^{\alpha'}E_{a_2}^{\beta'}E_{a_3}^{\gamma'}E_{a_4}^{\delta'}E_{a_5}^{\epsilon'}
   E_{a_6}^{\zeta'}
\right)
\mbox{\large $\otimes$} 
\nonumber  \\    
&\quad\mbox{\large $\otimes$} 
\left(\quad
Y\left(
e_{\alpha'} \otimes e_{\beta'}\otimes e_{\gamma'}
\otimes e_{\delta'} \otimes e_{\epsilon'}\otimes e_{\zeta'}
\right) \otimes
\right. 
\label{StrangeState}
\\  
& \quad\quad \otimes\left.
Y^{\rm tr}\left(
e^{b_1} \otimes e^{b_2}\otimes e^{b_3}\otimes e^{b_4}\otimes e^{b_5}\otimes e^{b_6}
\right)\quad
\right)
\nonumber
\end{align}
Here $E_{a_1}^{\alpha'}E_{a_2}^{\beta'}E_{a_3}^{\gamma'}E_{a_4}^{\delta'}E_{a_5}^{\epsilon'}
E_{a_6}^{\zeta'}$ is an element of the universal enveloping ${\cal U}{\bf g}$.
The corresponding vertex  is constant in AdS, because all the AdS (greek)
indices are contracted. It appears that there is no reason to discard this 
state. However observe that such states necessarily will have a nonzero 
contraction of the lower $a$ indices and the upper $b$ indices. It must be true 
that the vertex for (\ref{StrangeState}) is BRST-exact. 
But we have not checked it explicitly.

\section{BRST operator and parabolic induction}\label{sec:BRSTAndParabolicInduction}
In this an the following sections we will describe the construction of the
covariant vertex. We will start by calculating the action of the BRST
operator $Q$ in the induced representation.

\subsection{The bicomplex  $Q_L = Q_{L+} + Q_{L-}$}
\subsubsection{Anatomy of pure spinor}\label{sec:Anatomy}
Consider the ``left'' pure spinor $\lambda\in {\bf g}_3$. We get:
\begin{equation}
\lambda = \left( \begin{array}{cc} 0 & \lambda_+ \cr \lambda_- & 0 \end{array}\right)
\end{equation}
The condition that $\lambda\in {\bf g}_3$ implies:
\begin{equation}\label{LambdaMinusInTermsOfLambdaPlus}
(\lambda_-)^a_{\alpha} = i\; \omega^{aa'}(\lambda_+)_{a'}^{\alpha'} \omega_{\alpha'\alpha}
\end{equation}
where $\omega_{\alpha\beta}$, $\omega_{ab}$ is the symplectic form on ${\bf C}^{4|4}$ which defines the denominator 
subalgebra $sp(2)\subset sl(4)$ as explained in \cite{Roiban:2000yy}. In other words, the choice 
of $\omega$ up to multiplication by a number is equivalent to the choice of a point 
in $AdS_5\times S^5$. The purity condition implies that:
\begin{align}
\lambda^{\alpha}_a \omega^{ab} \lambda^{\beta}_b \simeq \omega^{\alpha\beta}
\label{FirstPurityCondition}
\\  
\lambda^{\alpha}_a \omega_{\alpha\beta} \lambda^{\beta}_b \simeq \omega_{ab}
\label{SecondPurityCondition}
\end{align}
In other words, pure spinors parametrize the group manifold of $Sp(2)$.
We also observe that (\ref{FirstPurityCondition}) $\Rightarrow$ (\ref{SecondPurityCondition}), therefore it is enough to impose only 5
pure spinor constraints (see {\it e.g.} Section 4.2 of \cite{Berkovits:2008ga}).

For brevity we will write $\lambda^{\alpha}_a = (\lambda_+)^{\alpha}_a$, {\it i.e.} drop the subindex $+$. 
Therefore the BRST operator $Q_L = \lambda^i t_i$ splits into the sum of the two 
anticommuting nilpotent operators:
\begin{align}
Q_L = &\; Q_{L+} + Q_{L-} 
\\    
& \; Q_{L+} = \; \lambda_+^i t_i 
\\    
& \; Q_{L-} = \; \lambda_-^i t_i 
\end{align}

\subsubsection{Notations $\cap$, $\cup$ and $||\ldots||$}
We will denote:
\begin{align}
X_a \omega^{ab} Y_b = X\cap Y \;,\;\;
X^{\alpha}\omega_{\alpha\beta}Y^{\beta} = X\cup Y \;,\;\;
\\  
X_{ab}\omega^{ba} = ||X||\;,\;\;
Y^{\alpha\beta}\omega_{\beta\alpha} = ||Y||
\end{align}

\subsubsection{The structure of $Q_{L+}$}
The action of $Q_{L+}$ on $\Phi$ follows from (\ref{ShiftOfThetaPlus}):
\begin{equation}
Q_{L+} \Phi^{\alpha_1\cdots \alpha_{2n}}_{a_1\cdots a_{2n}} (\theta_+) = 
\left(\lambda^{\alpha}_a {\partial\over\partial (\theta_+)^{\alpha}_a}\right) \;
\Phi^{\alpha_1\cdots \alpha_{2n}}_{a_1\cdots a_{2n}} (\theta_+) 
\end{equation}
Therefore:
\begin{itemize}
\item the action of $Q_{L+}$ coincides with the flat space zero mode BRST cohomology.
\end{itemize}
We will often abbreviate $\theta_+$ to $\theta$.

\subsubsection{The structure of $Q_{L-}$}
The expression for $Q_{L-}$ is somewhat more involved. From (\ref{ActionOfQMinus}) we get:
\begin{equation}\label{QLMinusPhi0}
\epsilon Q_{L-} \Phi = i\left[
 \theta {\cap} \epsilon \lambda \cup \Phi  
+  \Phi \cap \epsilon\lambda \cup \theta 
- \left((\theta\cap\epsilon\lambda\cup\theta){\partial\over\partial\theta}\right)\Phi
\right]
\end{equation}

\paragraph     {A redefinition of $Q$ and $\Phi$}
To make the formulas look better, we would prefer to get rid of the $i$ in 
the RHS of (\ref{QLMinusPhi}). This can be done by the redefinition:
\begin{align}
\Phi\mapsto & \exp\left[ {i\pi\over 4} (\#\theta)\right] \Phi
\\    
Q\mapsto    & \exp\left[ {i\pi\over 4} (\#\theta)\right] Q
\end{align}
where $\#\theta$ is the number of $\theta$'s in the expression. After this 
redefinition, we get:
\begin{equation}\label{QLMinusPhi}
\epsilon Q_{L-} \Phi = 
 \theta {\cap} \epsilon \lambda \cup \Phi  
+  \Phi \cap \epsilon\lambda \cup \theta 
- \left((\theta\cap\epsilon\lambda\cup\theta){\partial\over\partial\theta}\right)\Phi
\end{equation}
Assuming that $\Phi$ is even ({\it i.e.} contains even number of $\theta$'s) we get:
\begin{equation}\label{FormulaForQLMinus}
Q_{L-}\Phi = -\theta{\cap}\lambda \cup\Phi 
+ \Phi \cap \lambda{\cup}\theta +
\left((\theta\cap\lambda\cup\theta){\partial\over\partial\theta}\right)\Phi
\end{equation}
\remv{Check nilpotence:}\rem{photo/shot0014.png}

\subsection{The spectral sequence of $Q_{L+}+Q_{L-}$}\label{sec:SpectralSequence}
Let us look closer at the spectral sequence corresponding to the bicomplex
$Q_{L+} + Q_{L-}$. We will consider the filtration by the following degree:
\begin{equation}
\mbox{deg}(\Phi) \; = \; {1\over 2}\left[
   \mbox{number of $\theta$'s plus number of $\lambda$'s in $\Phi$}
\right]
\end{equation}
Notice that 
\begin{align}
\mbox{deg}(Q_{L+}\Phi) = & \; \mbox{deg}(\Phi)
\\    
\mbox{deg}(Q_{L-}\Phi) = & \; \mbox{deg}(\Phi) + 1
\end{align}
The first page of the spectral sequence for $Q_{L+} + Q_{L-}$ is:
\begin{equation}
E_1^{p,q} = \frac%
{\mbox{ker}\left(Q_{L+}:\; [\lambda^{p+q} \theta^{p-q}] \to [\lambda^{p+q+1}\theta^{p-q-1}] \right)}%
{\mbox{im}\left(Q_{L+}:\;  [\lambda^{p+q-1} \theta^{p-q+1}] \to [\lambda^{p+q}\theta^{p-q}] \right)}%
\end{equation}
In other words, $E_1^{p,q} = H(Q_{L+})|_{\lambda^{p+q}\theta^{p-q}}$. On the second page: 
\begin{equation}
E_2^{p,q} = H(Q_{L-}, H(Q_{L+}))|_{\lambda^{p+q}\theta^{p-q}}
\end{equation}
We will be calculating the differentials $d_1$ and $d_2$:
\begin{align}
d_1: &\;\; E_1^{p,q} \to E_1^{p+1,q}
\\    
d_2: &\;\; E_2^{p,q} \to E_2^{p+2,q-1}
\end{align}
Schematically:
\begin{align}
d_1: & \; ([\lambda^m\theta^n] + \ldots ) \to 
([\lambda^{m+1}\theta^{n+1}] + \ldots )
\\   
d_2: & \; ([\lambda^m\theta^n] + \ldots ) \to 
([\lambda^{m+1}\theta^{n+3}] + \ldots )
\end{align}

\subsection{Structure of $H(Q_{L+})$}
Let us denote: 
\begin{equation}
\leftsub{d}{\{}\lambda \stackrel{0}{\cup} \theta\}_c  
= \leftsub{d}{\{}\lambda \cup \theta\}_c 
- {1\over 4} \omega_{dc}||\lambda\cup\theta||  
\end{equation}
--- this is the ``$\omega$-traceless part'', {\it i.e.} $\omega^{dc}\leftsub{d}{\{}\lambda \stackrel{0}{\cup} \theta\}_c = 0$.

The cohomology group $H(Q_{L+})$ is generated by the following elements:
\begin{align}
W^{\alpha\beta} = &\; \leftsup{\alpha}{\{} \theta\stackrel{0}{\cap}\lambda \}^{\beta}
\\   
W_{ab} = &\; \leftsub{a}{\{} \theta\stackrel{0}{\cup}\lambda \}_b
\\    
W^{\alpha}_a = & \; 
\theta^{\alpha}\cap \lambda\cup \theta_a
\\   
Z^{\alpha}_{b} = &\; 
\theta^{\alpha}\cap\lambda\cup\theta\cap\lambda\stackrel{0}{\cup}\theta_b
\\    
Z^{\alpha\beta} = & \; 
\theta^{\alpha}\cap\lambda\cup\theta\cap\theta\cup\lambda\cap\theta^{\beta}
\\    
Z_{ab} = & \; 
\theta_a\cup\lambda\cap\theta\cup\theta\cap\lambda\cup\theta_b
\\   
S      = & \;  
|| \theta\cup\lambda\cap\theta\cup\theta\cap\lambda\cup\theta\cap\lambda\stackrel{0}{\cup}\theta ||
\end{align}
In the $\Gamma$-matrix notations:
\begin{align}
W^{\alpha\beta} & \mapsto (\theta\Gamma_{\urmbf{a}}\lambda) 
\label{WAlphaBeta}
\\   
W_{ab} &  \mapsto  (\theta\Gamma_{\urmbf{s}}\lambda) 
\label{WAB}
\\   
3W^{\alpha}_a  \simeq  ( \theta\cap\{\lambda\stackrel{0}{\cup}\theta\} +
\{\theta\stackrel{0}{\cap}\lambda\}\cup \theta) 
&  \mapsto \widehat{F} \Gamma_m\theta (\lambda\Gamma^m\theta) 
\\   
Z^{\alpha}_b \simeq  &  \mapsto 
(\lambda\Gamma^m\theta)(\lambda\Gamma^n\theta) \Gamma_{mn}\theta
\\    
Z^{\alpha\beta} \simeq  &  \mapsto
(\lambda\Gamma^m\theta)(\lambda\Gamma^n\theta) (\theta\Gamma_{\urmbf{a}mn}\theta)
\\    
Z_{ab} \simeq  &  \mapsto
(\lambda\Gamma^m\theta)(\lambda\Gamma^n\theta) (\theta\Gamma_{\urmbf{s}mn}\theta)
\\    
S &  \mapsto 
(\lambda\Gamma^k\theta)
(\lambda\Gamma^m\theta)(\lambda\Gamma^n\theta) (\theta\Gamma_{kmn}\theta)
\end{align}
In Eqs. (\ref{WAlphaBeta}) and (\ref{WAB}) the subindices $\;\urmbf{a}$ and $\;\urmbf{s}$ enumerate the tangent space 
to $AdS_5\times S^5$, namely $\;\urmbf{a}$ enumerates the tangent space to $AdS_5$ and $\;\urmbf{s}$ the 
tangent space to $S^5$. (Therefore both $\;\urmbf{a}$ and $\;\urmbf{s}$ run from 1 to 5.)

\subsection{Multiplication in $H(Q_{L+})$}
\paragraph     {Lemma}
Suppose $X_{dc}$ is an antisymmetric $\omega$-less rank 2 tensor 
({\it i.e.} $X_{dc} = - X_{cd}$ and $\omega^{dc}X_{dc} = 0$) and $\Psi_b$ is a rank 1 tensor. Then:
\begin{align}
\Psi_b X_{dc} = &
- {1\over 5} \Psi \cap X_{\bullet b} \omega_{dc} 
- {4\over 5} \Psi \cap X_{\bullet [d} \omega_{c]b} + P_{b[dc]}
\\    
\mbox{\small where}& \nonumber  
\\  
\omega^{bd}P_{b[dc]} = \omega^{bd}P_{c[bd]} = & \; 0
\end{align}
\remv{$ \omega^{bd}P_{b[dc]} = 0 $}\rem{ photo/shot0015.png }
Taking this an similar linear-algebraic identities into account, we derive 
the following multiplication table in $H(Q_{L+})$:
\begin{align}
W^{\alpha}_a \; W_{bc}  = &\; - {1\over 5} Z^{\alpha}_a \; \omega_{bc} 
- {4\over 5} Z^{\alpha}_{[b} \; \omega_{c]a}
\label{MultiplicationTLTxTL}\\    
W^{\alpha}_a W^{\beta}_b = &\; {1\over 4} \omega_{ab} Z^{\alpha\beta}
+ {1\over 4} \omega^{\alpha\beta} Z_{ab}
\\     
W^{\alpha}_a Z^{\beta}_b = &
\; {1\over 16} \omega_{ab}\omega^{\alpha\beta} S
\end{align}
The other products are zero.  

\subsection{The meta-symmetry}
The ``meta-symmetry'' is flipping all Greek letters with the corresponding
Latin letters, upper indices with lower indices, and $\cup$ with $\cap$. It is useful
to keep track of whether the expression is meta-odd or meta-even. Notice
that $Q_{L+}$ is meta-even, and $Q_{L-}$ is meta-odd.

\section{The (2,0) part}\label{sec:TwoZeroPart}
We will investigate the cohomology of $Q_L$ in the ghost numbers $(p,0)$ using 
the spectral sequence of $Q_{L+} + Q_{L-}$. We will find that the cohomology is 
trivial for all $p$.

\subsection{Cohomology of $Q_{L+}$}
Let us start by looking at the cohomology of $Q_{L+}$. 
There are classes of the types:
\begin{align}
 \lambda^0\theta^0 \;:\;
&\left(\omega^{\bullet\bullet}\otimes\omega_{\bullet\bullet}\right)^{\otimes (n+1)}
\label{LambdaZeroThetaZero}
\\  
 \lambda\theta \;:\;
&
\leftsup{\bullet}{\{}\theta\stackrel{0}\cap\lambda\}^{\bullet} \otimes
\left(\omega^{\bullet\bullet}\right)^{\otimes n} \otimes 
\left(\omega_{\bullet\bullet}\right)^{\otimes (n+1)} 
\pm \mbox{\small (meta-flip)}
\\  
& \mbox{\small (two classes of the type $\lambda\theta$,
one meta-odd and one meta-even)}
\nonumber
\\  
 \lambda^2\theta^4\;:\; &
\leftsub{\bullet\;}\theta
\cup\lambda\cap\theta\cup\theta\cap\lambda\cup\theta_{\bullet}
\otimes (\omega_{\bullet\bullet})^{\otimes n}
\otimes (\omega^{\bullet\bullet})^{\otimes (n+1)} \pm \mbox{\small (meta-flip)}
\\
& \mbox{\small (two classes of the type $\lambda^2\theta^4$,
one meta-odd and one meta-even)}
\nonumber
\\    
 \lambda^3\theta^5\;:\;
& ||\theta\cap \lambda\cup \theta \cap \theta \cup \lambda \cap \theta\cup \{\lambda\cap\theta\}||\;(\omega^{\bullet\bullet})^{\otimes (n+1)}\otimes
(\omega_{\bullet\bullet})^{\otimes (n+1)}
\end{align}
Notice that $\lambda^2\theta^3$ is missing because of the wrong quantum numbers.

\subsection{Higher differentials}

\begin{figure}[h!]
\centering
\includegraphics[scale=0.25]{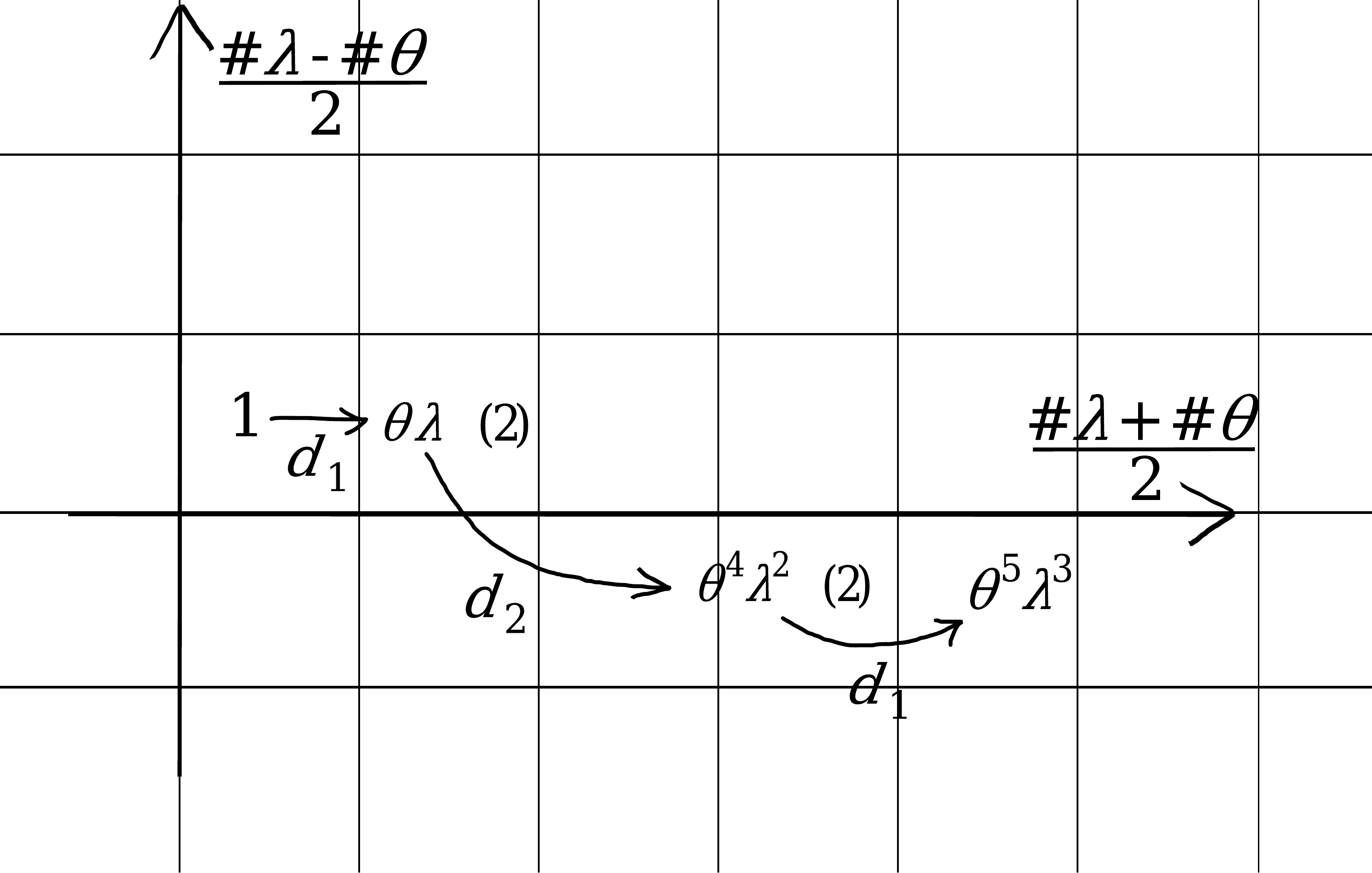}
\caption{Higher differentials on pages $E_1$ and $E_2$}
\end{figure}

\subsubsection{The meta-odd class of the type $\lambda\theta$ is cancelled by $d_1$ of $\lambda^0\theta^0$}
This follows immediately from (\ref{LambdaZeroThetaZero}) and (\ref{QLMinusPhi}).

\subsubsection{Two classes of the type $\lambda^2\theta^4$: $\Phi_0^+$ and $\Phi_0^-$}\label{sec:PhiZero}
There are two cohomology classes\footnote{Notice that $\lambda^2\theta^3$ does
not intertwine properly} of $Q_{L+}$. Let us consider {\it e.g.} $n=1$:
\begin{align}
(\Phi^{\pm}_0)^{\alpha\beta\gamma\delta}_{abcd} = &\;
{1\over 4} \;
\theta_{[a}\cup\lambda\cap\theta\cup
\theta\cap\lambda\cup\theta_{b]}\; 
\omega^{\alpha\beta} \; \omega^{\gamma\delta} \; \omega_{cd} \;\pm
\nonumber \\     
\pm &\; 
{1\over 4} \;
\theta^{[\alpha}\cap\lambda\cup\theta\cap
\theta\cup\lambda\cup\theta^{\beta]}\; 
\omega_{ab} \; \omega^{\gamma\delta} \; \omega_{cd} 
\label{PhiZeroAfterContraction}
\end{align}
In the notations of Section \ref{sec:SpectralSequence}, this means that $E_1^{3,-1}$ is two-dimensional, 
generated by $\Phi_0^{\pm}$. Notice that in $Q_{L+}$ cohomology $\Phi_0^+$ is equivalent to:
\begin{equation}
(\Phi_0)^{\alpha\beta\gamma\delta}_{abcd} = 
(\theta\cap\lambda\cup\theta)^{[\alpha}_{[a}\; 
(\theta\cap\lambda\cup\theta)^{\beta]}_{b]}\; 
\omega^{\gamma\delta} \; \omega_{cd} 
\end{equation}
It is easy to check that: 
\begin{equation}
Q_{L+}\Phi_0^+ = Q_{L+}\Phi_0^- = 0
\end{equation}
Next we have to calculate the action of $Q_{L-}$. It will turn out that $\Phi_0^-$ is 
not annihilated by $d_1$, and $\Phi_0^+$ is in the image of $d_2$. Therefore there is 
no cohomology in the ghost number $(2,0)$. 

\subsubsection{The class of the type $\lambda^3\theta^5$ is cancelled by $d_1$ of $\Phi_0^-$}
We will see that:
\begin{itemize}
\item $\Phi_0^+$ is annihilated by $d_1$
\item But $\Phi_0^-$ is not:
\begin{equation}
d_1\Phi_0^- = [\lambda^3\theta^5]
\end{equation}
Therefore $d_1\Phi_0^-$ cancels $||\theta\cap \lambda\cup \theta \cap \theta \cup \lambda \cap \theta\cup \{\lambda\cap\theta\}||\;\omega\cdots\omega\in E_1^{4,-1}$
\end{itemize}
Let us start with $\Phi_0^+$. Consider the case $n=1$, the general case is
completely analogous. We observe that $Q_{L-}\Phi_0^+$ is not literally zero:
\begin{equation}
\epsilon Q_{L-} \Phi^+_0 = (\theta\cap\lambda\cup\theta)\; \otimes\; 
(\theta\cap\lambda\cup\theta)\; \otimes \;
\left([\theta\cap\epsilon\lambda]\;\otimes\;\omega_{\bullet\bullet}
+ \omega^{\bullet\bullet} \;\otimes\;[\epsilon\lambda\cup\theta]\right)
\end{equation}
Up to $Q_{L+}$-exact terms this can be rewritten in the following way:
\begin{align}
(Q_{L-} \Phi^+_0)^{\alpha\beta\gamma\delta}_{abcd} = &  \theta^{\alpha}\cap\lambda\cup\theta_a\; \otimes\; 
\theta_b\cup\lambda\cap\theta^{\beta}
\left( 
  - \{\theta\cap\lambda\}^{\gamma\delta}\omega_{cd} 
  + \omega^{\gamma\delta}\{\theta\cup\lambda\}_{cd}
\right) +
\nonumber \\  
& + Q_{L+}(\ldots)
\end{align}
This is $Q_{L+}$-equivalent to:
\begin{align}
& {1\over 4}\theta^{[\alpha}\cap \lambda\cup \theta \cap \theta \cup \lambda \cap \theta^{\beta]}\;\omega_{ab}\;
(\{\theta\cap\lambda\}^{\gamma\delta}\omega_{cd} -\omega^{\gamma\delta}\{\theta\cup\lambda\}_{cd}) +
\label{BeforeFinalContraction}
\\ 
+ & \;{1\over 4}\theta_{[a}\cup\lambda\cap\theta\cup\theta\cap\lambda\cup\theta_{b]}\; \omega^{\alpha\beta}\;
(\{\theta\cap\lambda\}^{\gamma\delta}\omega_{cd} -\omega^{\gamma\delta}\{\theta\cup\lambda\}_{cd}) 
\end{align}
This is $Q_{L+}$-equivalent to:
\begin{align}
{1\over 4}
\theta^{[\alpha}\cap \lambda\cup \theta \cap \theta \cup \lambda \cap \theta^{\beta]}\;\omega_{ab}\;
\{\theta\cap\lambda\}^{\gamma\delta}\omega_{cd}  -
\nonumber \\   
- {1\over 4}
\theta_{[a}\cup\lambda\cap\theta\cup\theta\cap\lambda\cup\theta_{b]}\; \omega^{\alpha\beta}\;
\omega^{\gamma\delta}\{\theta\cup\lambda\}_{cd}
\label{QLMinusPhiLLTwoLines}
\end{align}
Now let us fuse $\theta\lambda\theta\theta\lambda\theta$ with $\{\theta\lambda\}$. We get:
\begin{align}
& {1\over 4}
\theta^{[\alpha}\cap \lambda\cup \theta \cap \theta \cup \lambda \cap \theta^{\beta]}\;\omega_{ab}\;
\{\theta\cap\lambda\}^{\gamma\delta}\omega_{cd}  -
\nonumber \\   
& - {1\over 4}
\theta_{[a}\cup\lambda\cap\theta\cup\theta\cap\lambda\cup\theta_{b]}\; \omega^{\alpha\beta}\;
\omega^{\gamma\delta}\{\theta\cup\lambda\}_{cd} =
\nonumber \\[5pt] 
= & \;
{1\over 16}
||\theta\cap \lambda\cup \theta \cap \theta \cup \lambda \cap \theta\cup \{\lambda\cap\theta\}||
\times
\nonumber \\  
& \times\;\left(
   \omega^{[\delta|\;[\alpha}\;\omega^{\beta]\;|\gamma]} \;\omega_{ab}\;\omega_{cd}  \;-\;
   \omega^{\alpha\beta}\omega^{\gamma\delta} \omega_{[d|\;[a}\; \omega_{b]\;|c]}
\right) \; + \; Q_{L+}(\ldots)
\label{FusionOfTLTTLTWithTL}
\end{align}
One can see that this vanishes after the symmetrization of 
$(ac)$ and $(bd)$. 

For $\Phi_0^-$ the relative sign of the two $\omega\omega\omega\omega$ terms is opposite 
compared to (\ref{FusionOfTLTTLTWithTL}), and therefore $d_1 \Phi_0^{-} \neq 0$.

\subsubsection{$\Phi_0^+$ is cancelled by $d_2$ of the meta-even class of the type $\lambda\theta$}\label{sec:D2E10}
We have so far proven that $\Phi_0^+$ is a nontrivial element of $E_2^{3,-1}$.
But we will now show that $\Phi_0^+$ is in the image of $d_2$:
\begin{equation}
d_2:\; E_2^{1,0} \to E_2^{3,-1}
\end{equation}
and therefore $\Phi_0^+$ does not survive on the next page $E_3^{\bullet\bullet}$.

Consider for example the case $n=1$. The $E_2^{1,0}$ is generated by $A^{\alpha\beta\gamma\delta}_{abcd}$:
\begin{align}
A = & \; \leftsup{\bullet}{\{}\theta\stackrel{0}\cap\lambda\}^{\bullet}
\; \left(\omega^{\bullet\bullet}\right)^{\otimes n} 
\; \left(\omega_{\bullet\bullet}\right)^{\otimes (n+1)}
\end{align}
We get:
\begin{equation}
\left(Q_{L-}A\right) =
\; (n+1) \; \leftsup{\bullet}{\{}\theta\stackrel{0}{\cap}\lambda\}^{\bullet}
\;\leftsub{\bullet}{\{}\theta\stackrel{0}{\cup}\lambda\}_{\bullet} \;
\left(\omega^{\bullet\bullet} \; \omega_{\bullet\bullet} \right)^{\otimes n}
\end{equation}
\rem{ photo/shot0004.png  }
This is equal to $Q_{L+} C$ where
\begin{align}
C= & 
   \;\left(
      \theta^{[\bullet}_{[\bullet} 
      \;\; \leftsup{\bullet]}{\{}\theta\cap\lambda\}\cup\theta_{\bullet]}
   + \theta^{[\bullet}_{[\bullet} \;\; 
   \theta^{\bullet]}\cap\{\lambda\cup\theta\}_{\bullet]} +
\right.
\label{QLPlusInvQLMinusA}\\    
& \left.  + {1\over 4} \omega^{\bullet\bullet}\;
\theta_{[\bullet}\cup\theta\cap\{\lambda\cup\theta\}_{\bullet]}
+ {1\over 4} \omega_{\bullet\bullet} \; \leftsup{[\bullet}{\{}\theta\cap\lambda\}\cup\theta\cap\theta^{\bullet]}
\right)
\;\left(\omega^{\bullet\bullet} \; \omega_{\bullet\bullet} \right)^{\otimes n}
\nonumber
\end{align}
(Notice that $\theta_{[a}\cup\{\theta\cap\lambda\}\cup\theta_{b]}$ is zero.) Let us now calculate $Q_{L-}C$, modulo
$\mbox{Im}(Q_{L+})$. It is useful to represent $C$ in the gamma-matrix notation:
\begin{equation}
C_{as} = (\theta\Gamma_{\urmbf{a}}\Gamma_{\urmbf{s}} \Gamma_m\theta) (\theta \Gamma^m\lambda) \otimes
\;\omega\cdots\omega
\end{equation}
where $\;\urmbf{a}$ enumerates the tangent space to $AdS_5$ and $\;\urmbf{s}$ the tangent space to 
$S^5$. (Therefore both $\;\urmbf{a}$ and $\;\urmbf{s}$ run from 1 to 5.) We have to act on this by 
$Q_{L-}$ and discard the terms which are $Q_{L+}$-exact. The $Q_{L+}$ cohomology $\theta^4\lambda^2$ 
has the following quantum numbers under $so(5)\oplus so(5)$: either vector under 
the first $so(5)$ and the scalar under the second $so(5)$ or vice versa. We can 
simply throw away everything else. In particular, when acting on $C$ with $Q_{L-}$
we can throw away the terms arizing from the action by the $\theta\lambda\theta{\partial\over\partial\theta}$ in $Q_{L-}$ 
--- see (\ref{FormulaForQLMinus}), because they do not change the $so(5)\oplus so(5)$ quantum numbers. 
The other terms are contractions with $\theta\cap\lambda$ and $\lambda\cup\theta$. We split: 
\begin{equation}
\theta\cap\lambda = - {1\over 4}||\theta\cap\lambda||\;\omega^{\bullet\bullet} +
 \theta\stackrel{0}{\cap}\lambda
\end{equation}
and similarly $\lambda\cup\theta$. The contraction with the first term $-{1\over 4}||\theta\cap\lambda||\;\omega^{\bullet\bullet}$ 
does not change the $so(5)\oplus so(5)$ quantum numbers, and therefore can be 
discarded. What remains is the rotation with the $\omega$-less part $\theta\stackrel{0}{\cap}\lambda$, which is
 $(\theta\Gamma_{\urmbf{a}}\lambda)$ in the $\Gamma$-matrix notations, and the $\theta\stackrel{0}{\cup}\lambda$, which is $(\theta\Gamma_{\urmbf{s}}\lambda)$. Let us 
look at this rotation from the following point of view. Consider $S^5$ embedded 
into ${\bf R}^6$, and $AdS_5$ into ${\bf R}^{2+4}$. Therefore $\omega$ represents a unit vector in ${\bf R}^6$,
and a unit vector in ${\bf R}^{2+4}$. What kind of an object is $C$ of (\ref{QLPlusInvQLMinusA})? It is a
product a traceless symmetric tensor $\left({\bf R}^{2+4}\right)^{\otimes (n+1)}_{\rm Symm,0}$ and a traceless symmetric 
tensor $\left({\bf R}^6\right)^{\otimes (n+1)}_{\rm Symm,0}$. Let us concentrate on the $S^5$ part of the rotation, 
{\it i.e.} on the lower (latin) indices of $C$. We have:
\begin{equation}
C = V\otimes \omega^{\otimes n} - \mbox{traces}
\end{equation}
where $V$ is the $S^5$ part of $(\theta\Gamma_{\urmbf{a}}\Gamma_{\urmbf{s}} \Gamma_m\theta)(\theta \Gamma^m\lambda)$. The vector $Y = (\theta\Gamma_{\urmbf{s}}\lambda)$ is
tangent to the $S^5$ at the point $\omega$. We need to rotate $C$ by the infinitesimal
rotation corresponding to the bivector $Y\wedge \omega$. The rotated tensor $(Y\wedge\omega).C$ 
is, again, a symmetric and traceless tensor of $so(6)$. It can be represented 
as a sum of a term proportional to $\omega^{\otimes (n+1)}$, and other terms which are 
orthogonal to $\omega^{\otimes (n+1)}$. We observe that the term proportional to $\omega^{\otimes (n+1)}$ 
in $(Y\wedge\omega).C$ equals to $-(Y,V) \;\omega^{\otimes (n+1)}$. Therefore: 
\begin{equation}
(Y,V)\neq 0 \;\mbox{ implies }\; (Y\wedge \omega).C \neq 0 
\end{equation}
The scalar product $(Y,V)$ where  $Y = (\theta\Gamma_{\urmbf{s}}\lambda)$ and $V = (\theta\Gamma_{\urmbf{a}}\Gamma_{\urmbf{s}} \Gamma_m\theta)(\theta \Gamma^m\lambda)$ 
equals\footnote{When the index $\urmbf{s}$ is contracted we assume the
summation over the {\em five} indices enumerating the tangent space ot $S^5$}:
\begin{equation}\label{ScalarProductYV}
(Y,V) = (\theta\Gamma^{\;\urmbf{s}}\lambda)
(\theta\Gamma_{\urmbf{a}}\Gamma_{\urmbf{s}} \Gamma_m\theta) (\theta \Gamma^m\lambda) 
\end{equation}
This is a nontrivial class of the $Q_{L+}$ cohomology, Notice that the index $\urmbf{a}$ 
remained from the AdS part of $C$, so this is $(Q_{L-}C)_{\;\urmbf{a}}$. The $(Q_{L-}C)_{\;\urmbf{s}}$ is 
given by a similar expression. The expression (\ref{ScalarProductYV}) is $Q_{L+}$-equivalent to:
\begin{equation}
\theta^{[\alpha}\cap\lambda\cup\theta\cap\theta\cup\lambda\cap\theta^{\beta]}
\end{equation}
This means that $d_2 A$ is in fact identified with the $\Phi_0$, as given by 
Eq. (\ref{PhiZeroAfterContraction}):
\begin{align}
&\;
{1\over 4} \;
\theta_{[\bullet}\cup\lambda\cap\theta\cup
\theta\cap\lambda\cup\theta_{\bullet]}
\; \left(\omega^{\bullet\bullet}\right)^{\otimes (n+1)} 
\; \left(\omega_{\bullet\bullet} \right)^{\otimes n} +
\\     
+ &\; 
{1\over 4} \;
\theta^{[\bullet}\cap\lambda\cup\theta\cap
\theta\cup\lambda\cap\theta^{\bullet]}\; 
\; \left(\omega^{\bullet\bullet}\right)^{\otimes n} 
\; \left(\omega_{\bullet\bullet}\right)^{\otimes (n+1)}
\end{align}
(The relative sign of the two terms is such that this is meta-even,
since (\ref{QLPlusInvQLMinusA}) is meta-odd.)

\subsubsection{Conclusion: the cohomology of $Q_{L}$ in expressions which do not
contain $\mu$ is trivial}
Indeed, we have seen that all the cohomology classes of $Q_{L+}$ cancel
when we correct $Q_{L+} \to Q_L = Q_{L+} + Q_{L-}$. This shows that
the (2,0) part of the vertex can always be gauged away. (And the same is true
about the (0,2) part.)

\section{The (1,1) part}\label{sec:OneOnePart}
\subsection{Notations and summary}
In this section we will show that $Q = Q_L + Q_R$ has nontrivial
cohomology in the ghost number (1,1). 

We will introduce the right pure spinor which will be
denoted $\mu$. As in Section \ref{sec:Anatomy} we will write:
\begin{equation}
\mu = \left( 
   \begin{array}{cc} 0 & \mu_+ \cr \mu_- & 0 \end{array}\right)
\end{equation}
but now there is a minus sign in the expression relating $\mu_-$ to $\mu_+$,
compared to Eq. (\ref{LambdaMinusInTermsOfLambdaPlus}):
\begin{equation}\label{MuMinusInTermsOfMuPlus}
(\mu_-)^a_{\alpha} = - i\omega^{aa'}(\mu_+)_{a'}^{\alpha'} \omega_{\alpha'\alpha}
\end{equation}
This is because $\mu\in {\bf g}_1$ while $\lambda\in{\bf g}_3$. This leads to the overall
minus sign in the action of $Q_{R-}$, compared to Eq. (\ref{QLMinusPhi}):
\begin{equation}\label{QRMinusPhi}
\epsilon Q_{R-} \Phi = 
- \theta {\cap} \epsilon \mu \cup \Phi  
-  \Phi \cap \epsilon\mu \cup \theta 
+ \left((\theta\cap\epsilon\mu\cup\theta){\partial\over\partial\theta}\right)\Phi
\end{equation}
\subsection{The cohomology of $Q_{L+}$ on expressions linear in $\lambda$ and $\mu$}
The cohomology of $Q_{L+}$ in the sector of functions linear in both $\lambda$ and $\mu$ 
is generated by the following expressions:
\subsubsection{Two-trace combination}
This is the part with the maximal number of $\omega$'s:
\begin{equation}
\stackrel{\rm tr}{\Psi}(\theta) = 
||\mu \cap \theta\cup\lambda\cap\theta || 
\; (\omega_{\bullet\bullet}\otimes \omega^{\bullet\bullet})^{\otimes(n+1)}
\end{equation}
This is meta-odd.
In the Gamma-matrix notations this corresponds to $(\mu\Gamma_m\theta)(\lambda\Gamma^m\theta)$. 
\subsubsection{One-trace combinations}
There are two of them:
\begin{align}
\stackrel{\rm dn}{\Psi}(\theta) = & \; 
\left(
   \mu_{[\bullet} \cup \theta\cap\lambda\cup\theta_{\bullet]} -
   {1\over 4}\omega_{\bullet\bullet} ||\mu \cup \theta\cap\lambda\cup\theta||
\right)
(\omega_{\bullet\bullet})^{\otimes n} (\omega^{\bullet\bullet})^{\otimes (n+1)}
\\   
\stackrel{\rm up}{\Psi}(\theta) = & \; 
\left(
   \mu^{[\bullet} \cap \theta\cup\lambda\cap\theta^{\bullet]} -
   {1\over 4}\omega^{\bullet\bullet} ||\mu \cap \theta\cup\lambda\cap\theta||
\right)
(\omega_{\bullet\bullet})^{\otimes (n+1)} (\omega^{\bullet\bullet})^{\otimes n}
\end{align}
They are interchanged by the meta-symmetry:
\begin{equation}
\mbox{meta}(\stackrel{\rm dn}{\Psi}) = \stackrel{\rm up}{\Psi}
\;\;,\;\;\;
\mbox{meta}(\stackrel{\rm up}{\Psi}) = \stackrel{\rm dn}{\Psi}
\end{equation}
In the Gamma-matrix notations they correspond to $(\mu\widehat{F} \Gamma_{\urmbf{s}}\Gamma_m\theta)(\theta\Gamma^m\lambda)$ and
$(\mu\widehat{F} \Gamma_{\urmbf{a}}\Gamma_m\theta)(\theta\Gamma^m\lambda)$.
\subsubsection{Double-traceless combination}
The traceless combination is:
\begin{align}
& \stackrel{0}{\Psi}(\theta) = \; \left(
\mu^{[\bullet}_{[\bullet} \; \theta^{\bullet]}\cap\lambda\cup\theta_{\bullet]} 
\right.
\nonumber \\    
& 
+ \; {1\over 4} 
\; \mu_{[\bullet} \cup \theta\cap\lambda\cup\theta_{\bullet]} 
\; \omega^{\bullet\bullet} 
- \; {1\over 4}
\; \mu^{[\bullet} \cap \theta\cup\lambda\cap\theta^{\bullet]} 
\;\omega_{\bullet\bullet} \; +
\nonumber \\   
& \left. 
+ \; {1\over 16}\;\omega_{\bullet\bullet}\; \omega^{\bullet\bullet}\;
||\mu \cap \theta\cup\lambda\cap\theta ||
\right) \otimes
\left(\omega^{\bullet\bullet}\otimes\omega_{\bullet\bullet}\right)^{\otimes n}
\end{align}
This is meta-odd. In the Gamma-matrix notations this is $(\mu\Gamma_{\urmbf{a}}\Gamma_{\urmbf{s}}\Gamma_m\theta)(\theta\Gamma^m\lambda)$.

\paragraph     {What happens when we act by $Q_{R+}$}
It turns out that both $Q_{R+}\stackrel{\rm up}{\Psi}$ and $Q_{R+}\stackrel{\rm dn}{\Psi}$ are nontrivial in the 
$Q_{L+}$-cohomology, while $Q_{R+}\stackrel{\rm tr}{\Psi}=0$ and $Q_{R+}\stackrel{0}{\Psi}$ is trivial in the 
$Q_{L+}$-cohomology. 
\remv{Psiup and Psidn are not annihilated by QR}\rem{ photo/100_0632.JPG }

\paragraph     {What happens when we act by $Q_{L-}$}
There is only one obstacle at $[\mu\lambda^2\theta^4]$ for constructing the full $Q_L$-closed 
expression. Therefore there is a linear combination of $\stackrel{0}{\Psi}$ and $\stackrel{\rm tr}{\Psi}$ that can be 
extended to the full $Q_L$-closed expression.

\subsection{$Q_{L-}$ on $\mu\otimes\theta\lambda\theta\otimes\mbox{$\omega$'s}$}
Let us consider the following expression:
\begin{align}
\Psi(\theta) = & \;
\mu^{[\bullet}_{[\bullet} \; \theta^{\bullet]}\cap\lambda\cup\theta_{\bullet]} 
\otimes \left(\omega^{\bullet\bullet}\right)^{\otimes n}
\otimes \left(\omega_{\bullet\bullet}\right)^{\otimes n}
\end{align}
Notice that this is meta-odd. Therefore $Q_{L-}\Psi$ is meta-even. This, in 
particular, implies, that when we calculate the action of $Q_{L-}$ on $\Psi$, the 
sum of terms arizing from the action of  $Q_{L-}$ on $\omega$ are all $Q_{L+}$-exact.
Indeed, these terms would have the form $\mu\otimes (\theta\lambda\theta)  \otimes Q_{L-}(\cup\otimes\cap)$. The 
fusion of $\theta\lambda\theta$ and $Q_{L-}(\cup\otimes\cap)$ produces a meta-even expression. But the 
cohomology obstacle, which is the cohomology  class of $Q_{L+}$, is of the form 
$\theta\cap\lambda\cup\theta\cap\lambda\cup\theta$, and is meta-odd. Therefore, we can neglect the terms 
which appear when $Q_{L-}$ acts on the $\omega$s. 

Also observe that $Q_{L-} \theta^{\bullet}\cap \lambda\cup\theta_{\bullet} = 0$. Therefore
the only nontrivial contribution arizes when $Q_{L-}$ acts on $\mu$:
\begin{align}
\left(
   \mu^{[\alpha}\cap\lambda\cup\theta_{[a} 
- \theta^{[\alpha}\cap\lambda\cup\mu_{[a}
\right)\; \theta^{\beta]}\cap\lambda\cup\theta_{b]} 
\end{align}
This is meta-even, and can be rewritten modulo $Q_{L+}$-exact terms as follows:
\begin{align}
{1\over 2}
\; \theta^{[\alpha}\cap\lambda\cup\theta_{[a} \;
\leftsub{b]}{\{} \theta\cup\lambda \}\cap\mu^{\beta]}  
+{1\over 2}
\; \theta_{[a}\cup\lambda\cap\theta^{[\alpha} \;
\leftsup{\beta]}{\{} \theta\cap\lambda \}\cup\mu_{b]}
\end{align}
Given the multiplication rule (\ref{MultiplicationTLTxTL}), this is equivalent in the
$Q_{L+}$-cohomology to the following expression:
\begin{align}
-{1\over 10} &
\; \theta^{[\alpha}\cap\lambda\cup\theta\cap
\{ \theta\stackrel{0}{\cup}\lambda \}\cap\mu^{\beta]}\;\omega_{ab}\;    
-{2\over 5} 
\; \theta^{[\alpha}\cap\lambda\cup\theta\cap
\{ \theta\stackrel{0}{\cup}\lambda \}_{[a}\;\mu_{b]}^{\beta]}\; -
\nonumber \\    
-{1\over 10} &
\; \theta_{[a}\cup\lambda\cap\theta\cup
\{ \theta\stackrel{0}{\cap}\lambda \}\cup\mu_{b]}\;\omega^{\alpha\beta}\; 
-{2\over 5} 
\; \theta_{[a}\cup\lambda\cap\theta\cup
\{ \theta\stackrel{0}{\cap}\lambda \}^{[\alpha}\;\mu_{b]}^{\beta]}\;
\end{align}
Notice that the two terms which have both indices of $\mu$ uncontracted cancel 
each other modulo $Q_{L+}$-exact terms. Therefore:
\begin{align}
Q_{L-}&\left(\mu^{[\bullet}_{[\bullet} \; 
\theta^{\bullet]}\cap\lambda\cup\theta_{\bullet]} 
\otimes 
(\omega^{\bullet\bullet})^{\otimes n}\otimes(\omega_{\bullet\bullet})^{\otimes n}
\right) =
\nonumber \\   
= &\; -{1\over 10}
\; \theta^{[\bullet}\cap\lambda\cup\theta\cap
\{ \theta\stackrel{0}{\cup}\lambda \}\cap\mu^{\bullet]}\otimes
(\omega^{\bullet\bullet})^{\otimes n}\otimes(\omega_{\bullet\bullet})^{\otimes (n+1)}
\;\; -
\nonumber \\    
& \; -{1\over 10}
\; \theta_{[\bullet}\cup\lambda\cap\theta\cup
\{ \theta\stackrel{0}{\cap}\lambda \}\cup\mu_{\bullet]}\otimes
(\omega^{\bullet\bullet})^{\otimes (n+1)}\otimes(\omega_{\bullet\bullet})^{\otimes n}
\;\; + 
\nonumber\\  
& + \; Q_{L+}\mbox{(smth)}
\label{QOnMuOtimesTLT}
\end{align}

\paragraph     {Comment: relating $Q_{L-}(\mu\otimes\theta\lambda\theta)$ to $Q_{R+}(\theta\lambda\theta\otimes\theta\lambda\theta)$}
Since we have shown that we can neglect the terms where $Q_{L-}$ is 
hitting $\omega$'s,  we get:
\begin{align}
& \; Q_{R+}\; \left(
   \theta^{[\alpha}\cap\lambda\cup \theta_{[a}\;\theta_{b]}\cup\lambda\cap\theta^{\beta]} 
   \; \omega^{\gamma\delta}\;\omega_{cd} 
\right) =
\nonumber \\   
= &\; Q_{L-}\;\left(
   \mu^{[\alpha}_{[a}\;\theta_{b]}\cup\lambda\cap\theta^{\beta]} 
   \; \omega^{\gamma\delta}\;\omega_{cd} 
\right)  
\end{align}

\subsection{$Q_{L-}$ on $\theta\lambda\theta\mu\otimes\mbox{$\omega$'s}$}
Let us now calculate this:
\begin{equation}
Q_{L-}\left(
   \theta_{[a}\cup\lambda\cap\theta\cup\mu_{b]} \;\; \omega^{\alpha\beta}
   \;\; \omega^{\gamma_1\delta_1}\cdots\omega^{\gamma_{n}\delta_{n}}
   \;\; \omega_{c_1d_1}\cdots\omega_{c_{n}d_{n}}\;
\right)
\end{equation}

\paragraph     {Comment on notations:} 
It turns out that the structure of the formulas has some regular dependence 
on $n$. It is enough to consider the case $n=1$, and then put the 
coefficient $n$ where necessary. To save space we will replace:
\begin{equation}
\omega^{\gamma_1\delta_1}\cdots\omega^{\gamma_{n}\delta_{n}} \;\; 
\omega_{c_1d_1}\cdots\omega_{c_{n}d_{n}}\; \longrightarrow \; \omega^{\gamma\delta}
\omega_{cd}
\end{equation}
but keep track of the coefficient $n$. We have:
\begin{align}
Q_{L-}\left(\theta_{[a}\cup\lambda\cap\theta\cup\mu_{b]} \;\; \omega_{cd} \; 
\omega^{\alpha\beta} \; \omega^{\gamma\delta}\right) = 
\\    
= \theta_{[a}\cup\lambda\cap\theta\cup\lambda\cap\theta\cup\mu_{b]}\;\;\omega_{cd}\;
\omega^{\alpha\beta} \; \omega^{\gamma\delta} \; + 
\\    
+ \; \theta_{[a}\cup\lambda\cap\theta\cup\mu\cap\lambda\cup\theta_{b]}\;\;
\omega_{cd}\; \omega^{\alpha\beta} \; \omega^{\gamma\delta} +
\\   
+ \; n \;
\theta_{[a}\cup\lambda\cap\theta\cup\mu_{b]} 
\;\;\omega^{\alpha\beta} \; \omega^{\gamma\delta}\{\lambda_c\cup\theta_d\} -
\\   
- \; (n+1) \;
\theta_{[a}\cup\lambda\cap\theta\cup\mu_{b]} \;\; \omega_{cd} \; 
\{\lambda^{\alpha}\cap\theta^{\beta}\}\; \omega^{\gamma\delta}
\end{align}
We  use the multiplication rule (\ref{MultiplicationTLTxTL}) to transform this expression. 
After a calculation, we get up to $Q_{L+}$-exact terms:
\begin{align}
Q_{L-} & \left( 
   \theta_{[a}\cup\lambda\cap\theta\cup\mu_{b]} 
   \;\; \omega_{cd} \; \omega^{\alpha\beta} \; \omega^{\gamma\delta} -
\right. \\   
& \left. 
   - \theta^{[\alpha}\cap\lambda\cup\theta\cap\mu^{\beta]}\;\omega^{\gamma\delta}\;
   \omega_{ab}\;\omega_{cd} 
\right) = 
\label{QOnTLTM}   \\    
= & -{1\over 5}\;(4n+3)\; 
\left(\;\;\mu_{[a}\cup\theta\cap\lambda\cup\theta\cap
\lambda\stackrel{0}{\cup}\theta_{b]} 
\;\;\omega_{cd}\;\omega^{\alpha\beta} \; \omega^{\gamma\delta} + 
\right.
\\
& \phantom{-{1\over 5}\;(4n+3)\;\;\;} \left.
+ \mu^{[\alpha}\cap\theta\cup\lambda\cap\theta\cup
\lambda\stackrel{0}{\cap}\theta^{\beta]} 
\;\;\omega^{\gamma\delta}\;\omega_{ab} \; \omega_{cd}\;\;\right)
\end{align}
Notice that the meta-odd pieces cancelled, and we are left on the RHS 
with the meta-even expression; this is because the LHS we have $Q_{L-}$ of the
meta-odd expression. 

\subsubsection{Acting on the double trace}
Again, the dependence on $n$ is regular. We consider first the case $n=0$:
\begin{align}
& Q_{L-}\left( 
||\mu \cap \theta\cup\lambda\cap\theta ||
\; \omega^{\alpha\beta}
\;\omega_{ab}
\right) =
\nonumber \\[5pt]    
= & \;||\mu \cap \theta\cup\lambda\cap\theta ||
\left(
   \leftsub{a}{\{}\lambda\cup\theta\}_b \; \omega^{\alpha\beta} 
   \;-\; \omega_{ab}\; \leftsup{\alpha}{\{}\lambda\cap\theta\}^{\beta}
\right)   =
\nonumber \\[5pt]   
= & -\; ||\mu\cup\theta\cap\lambda\cup\theta ||
\leftsub{a}{\{}\lambda\cup\theta\}_b 
\; \omega^{\alpha\beta}  \; -
\nonumber \\   
& -\; ||\mu \cap \theta\cup\lambda\cap\theta ||
\omega_{ab}\; \leftsup{\alpha}{\{}\lambda\cap\theta\}^{\beta}\;
= 
\nonumber \\[5pt]  
= &  \; {4\over 5} \;
\mu_{[b}\cup\theta\cap\lambda\cup\theta\cap\{\lambda\stackrel{0}{\cup}\theta\}_{a]}
\; \omega^{\alpha\beta} \;  + 
 \\     
& + \; {4\over 5} \;
\mu^{[\beta}\cap\theta\cup\lambda\cap\theta\cup\{\lambda\stackrel{0}{\cap}\theta\}^{\alpha]}
\;\omega_{ab}
\nonumber
\end{align}
\remv{Explanation:}\rem{photo/shot0017.png}
For general $n$:
\begin{align}
& Q_{L-}\left( 
||\mu \cap \theta\cup\lambda\cap\theta ||
\; (\omega^{\bullet\bullet})^{\otimes (n+1)}
\;(\omega_{\bullet\bullet})^{\otimes (n+1)}
\right) =
\nonumber \\[5pt]    
= &-  \; {4\over 5}(n+1) \;
\mu_{[\bullet}\cup\theta\cap\lambda\cup\theta\cap\{\lambda\stackrel{0}{\cup}\theta\}_{\bullet]}
\otimes (\omega_{\bullet\bullet})^{\otimes n}
\otimes (\omega^{\bullet\bullet})^{\otimes (n+1)} \;  -
\label{QOnDoubleTrace0} \\     
& - \; {4\over 5}(n+1) \;
\mu^{[\bullet}\cap\theta\cup\lambda\cap\theta\cup\{\lambda\stackrel{0}{\cap}\theta\}^{\bullet]}
\otimes (\omega^{\bullet\bullet})^{\otimes n}
\otimes (\omega_{\bullet\bullet})^{\otimes (n+1)} 
\nonumber
\end{align}
Let us denote:
\begin{equation}\label{PsiTrace}
\stackrel{\rm tr}{\Psi}(\theta) = 
||\mu \cap \theta\cup\lambda\cap\theta || 
\;
\left(\omega^{\bullet\bullet}\right)^{\otimes (n+1)}
\otimes
\left(\omega_{\bullet\bullet}\right)^{\otimes (n+1)}
\end{equation}

\subsubsection{Acting on the traceless combination with $Q_{L-}$}
The traceless combination is:
\begin{align}
& \stackrel{0}{\Psi}(\theta)^{\alpha\beta}_{ab} = \;
\mu^{[\alpha}_{[a} \; \theta^{\beta]}\cap\lambda\cup\theta_{b]} 
\\    
& 
+ \; {1\over 4} 
\; \mu_{[a} \cup \theta\cap\lambda\cup\theta_{b]} 
\; \omega^{\alpha\beta} 
- \; {1\over 4}
\; \mu^{[\alpha} \cap \theta\cup\lambda\cap\theta^{\beta]} 
\;\omega_{ab} \; +
\\   
& + \; {1\over 16}\;\omega_{ab}\; \omega^{\alpha\beta}\;
||\mu \cap \theta\cup\lambda\cap\theta ||
\end{align}
Summarizing Eqs.  (\ref{QOnMuOtimesTLT}), (\ref{QOnTLTM}) and (\ref{QOnDoubleTrace0}), we get:
\remv{Calculation:}\rem{\small
\begin{align}
Q_{L-}\stackrel{0}{\Psi} =&\;
-\; {1\over 10} \;
\theta^{[\alpha}\cap\lambda\cup\theta\cap\{\theta\stackrel{0}{\cup}\lambda\}\cap\mu^{\beta]} \;\omega_{ab} -
\\     
& 
- \; {1\over 10} \; 
\theta_{[a}\cup\lambda\cap\theta\cup
\{ \theta\stackrel{0}{\cap}\lambda \}\cup\mu_{b]}\;\omega^{\alpha\beta}\; +
\\   
& 
+ \; {1\over 20}\;(4n+3)\; 
\left(\;\;\mu_{[a}\cup\theta\cap\lambda\cup\theta\cap
\lambda\stackrel{0}{\cup}\theta_{b]} 
\;\omega^{\alpha\beta} \; + 
\right.
\\
& \phantom{-{1\over 20}\;(4n+3)\;\;\;} \left.
+\; \mu^{[\alpha}\cap\theta\cup\lambda\cap\theta\cup
\lambda\stackrel{0}{\cap}\theta^{\beta]} 
\;\omega_{ab} \;\right) \; -
\\   
& - \; {1\over 20}(n+1) \;
\mu_{[a}\cup\theta\cap\lambda\cup\theta\cap\{\lambda\stackrel{0}{\cup}\theta\}_{b]}
\; \omega^{\alpha\beta}\;  -
\\     
& - \; {1\over 20}(n+1) \;
\mu^{[\alpha}\cap\theta\cup\lambda\cap\theta\cup\{\lambda\stackrel{0}{\cap}\theta\}^{\beta]}
\;\omega_{ab}\;
\end{align}
This sums up to:
}
\begin{align}
Q_{L-}\stackrel{0}{\Psi} =\;
{2n+5\over 20}&\;\left(\mu_{[a}\cup\theta\cap\lambda\cup\theta\cap
\lambda\stackrel{0}{\cup}\theta_{b]} 
\;\omega^{\alpha\beta} \; \right. +
\\   
& + \left.
\mu^{[\alpha}\cap\theta\cup\lambda\cap\theta\cup
\lambda\stackrel{0}{\cap}\theta^{\beta]}\;\omega_{ab}
\right) 
\end{align}
\paragraph     {Expression in the kernel of $Q_{L-}$}
Therefore, let us define:
\begin{equation}
\Psi_{\rm complete} = {20\over 2n+5} \stackrel{0}{\Psi} 
+ {5\over 8n + 8} \stackrel{\rm tr}{\Psi} 
\end{equation}
We get:
\begin{equation}
Q_{L-} \Psi_{\rm complete} = Q_{L+} [\mu\lambda\theta^4]
\end{equation}
This means that we can redefine $\Psi_{\rm complete}$ by adding to it the
terms of the order $\theta^4$ and higher so that:
\begin{equation}
Q_{L}\Psi_{\rm complete}  = 0
\end{equation}

\subsection{Acting with $Q_{R+}$}
\subsubsection{Further adjustment of $\Psi_{\rm complete}$} 
The so constructed $\Psi_{\rm complete}$ is not annihilated by $Q_{R+}$. We will
now modify  $\Psi_{\rm complete}$ by adding to it a $Q_L$-exact expressions, so that 
the leading term of the modified $\Psi_{\rm complete}$ is annihilated by $Q_{R+}$.

We will start with the following modification:
\begin{align}
\Psi_{\rm complete} \mapsto & \; \Psi_{\rm complete} +
{20\over 2n+5}\times {1\over 3}\;Q_L\left( 
\mu^{[\alpha}_{[a} \; \theta^{\beta]}\cap\theta\cup\theta_{b]} \; +
\right. \\    
& 
+ \; {1\over 4} 
\; \mu_{[a} \cup \theta\cap\theta\cup\theta_{b]} 
\; \omega^{\alpha\beta} 
- \; {1\over 4}
\; \mu^{[\alpha} \cap \theta\cup\theta\cap\theta^{\beta]} 
\;\omega_{ab} \; +
\\   
&\left. + \; {1\over 16}\;\omega_{ab}\; \omega^{\alpha\beta}\;
||\mu \cap \theta\cup\theta\cap\theta || \right)
\end{align}
This only changes in the leading expression  of $\Psi_{\rm complete}$
is in the $\omega$-less part $\stackrel{0}{\Psi}$; the leading $\omega$-less part of the modified
 $\Psi_{\rm complete}$ is now the following:
\begin{align}
 \stackrel{0}{\Psi}(\theta)^{\alpha\beta}_{ab} = & \;
{20\over 2n+5}\times {1\over 3}\;\left(\mu^{[\alpha}_{[a} \; 
\leftsup{\beta]}{\{}\theta\stackrel{0}{\cap}\lambda\}\cup\theta_{b]} 
+
\mu^{[\alpha}_{[a} \; 
\theta^{\beta]}\cap\{\lambda\stackrel{0}{\cup}\theta \}_{b]}
\right)
+   
\nonumber \\  
& + \mbox{(subtraction of $\omega$-contractions)}
\end{align}
\rem{ photo/shot0005.png }
Observe that in the $\Gamma$-matrix notations this modified $\stackrel{0}{\Psi}$ has the 
following form:
\begin{equation}
\stackrel{0}{\Psi}_{\urmbf{as}} \simeq \left(\mu \Gamma_{\urmbf{a}}\Gamma_{\urmbf{s}} \Gamma_m \theta\right) 
\left( \lambda\Gamma^m \theta \right)
\end{equation}
where $\simeq$ means ``proportional to''. \rem{ photo/shot0006.png } 
(The index notations is as explained after Eqs. (\ref{WAlphaBeta}) and (\ref{WAB}).) We get: 
\begin{align}
& Q_{R+}\left(\mu \Gamma_{\urmbf{a}}\Gamma_{\urmbf{s}} \Gamma_m \theta\right) 
\left( \lambda\Gamma^m \theta \right)
\; = \;
-2\;(\mu\Gamma_{\urmbf{a}}\theta) (\lambda\Gamma_{\urmbf{s}}\mu)
+ 2\;(\mu\Gamma_{\urmbf{s}}\theta)(\lambda\Gamma_{\urmbf{a}}\mu) =
\nonumber
\\    
= &\; 2 Q_{L+} (\mu\Gamma_{\urmbf{a}}\theta)\;(\mu\Gamma_{\urmbf{s}}\theta) =
{1\over 2}
Q_{L+}Q_{R+} (\theta\Gamma_{\urmbf{a}}\Gamma_{\urmbf{s}}\Gamma_m\theta) (\mu\Gamma^m\theta)
\end{align}
\rem{ photo/shot0007.png }This implies:
\begin{align}
Q_{R+}\left(\left(\mu \Gamma_{\urmbf{a}}\Gamma_{\urmbf{s}} \Gamma_m \theta\right) 
(\lambda\Gamma^m\theta)
+ {1\over 2}Q_{L+}(\theta\Gamma_{\urmbf{a}}\Gamma_{\urmbf{s}}\Gamma_m\theta) (\mu\Gamma^m\theta)\right) = 0
\end{align}
The properties of the expression 
\begin{equation}\label{LeadingTracelessTermInPsi}
\Psi_{\urmbf{as}}^{0,\rm new}
=\left(\mu \Gamma_{\urmbf{a}}\Gamma_{\urmbf{s}} \Gamma_m \theta\right) 
(\lambda\Gamma^m\theta)
+ {1\over 2}Q_{L+}(\theta\Gamma_{\urmbf{a}}\Gamma_{\urmbf{s}}\Gamma_m\theta) (\mu\Gamma^m\theta)
\end{equation}
are studied in Appendix \ref{sec:GammaMatrixExpressions}, where it is shown that this expression
is symmetric under the exchange of $\lambda$ and $\mu$.

Let us  modify $\Psi_{\rm complete}$ once more by adding to it 
the $Q_L$-exact expression in the following way:
\begin{align}
\Psi_{\rm complete} = & \; \Psi_{\rm complete} \; +
\nonumber \\
&\; +
{20\over 2n+5}\times {1\over 6}\;Q_L\left(\theta^{[\alpha}_{[a} \; 
\leftsup{\beta]}{\{}\theta\stackrel{0}{\cap}\mu\}\cup\theta_{b]} 
+
\theta^{[\alpha}_{[a} \; 
\theta^{\beta]}\cap\{\mu\stackrel{0}{\cup}\theta \}_{b]}
\right. -
\nonumber \\    
&\;\left.\phantom{{20\over 2n+5}\times {1\over 6}Q_L\;\;\;\;\;\;}
-\mbox{$\omega$-traces}\right)
\end{align}
The leading term of the so modified $\Psi_{\rm complete}$ is in the $\Gamma$-matrix notations 
proportional to (\ref{LeadingTracelessTermInPsi}). Therefore the $Q_{R+}$ on the leading term of the 
modified $\Psi_{\rm complete}$ is zero. 

Now we have:
\begin{align}
 Q_R\Psi_{\rm complete}  = & \; [\mu^2\lambda\theta^3]
\\  
 Q_LQ_R\Psi_{\rm complete}  = &  \; 0
\end{align}
Therefore exists $\Phi = [\mu^2\theta^4]+\ldots$:
\begin{equation}
Q_R\Psi_{\rm complete}  + Q_L\Phi = 0
\end{equation}
Observe that $Q_LQ_R\Phi=0$, therefore $Q_R\Phi = 0$, therefore we find
a BRST-closed expression:
\begin{equation}\label{VIsInKernel}
(Q_L+Q_R)\;\left(\Psi_{\rm complete}  + \Phi\right) = 0
\end{equation}
Let us denote:
\begin{equation}
v(\lambda,\mu) = \Psi_{\rm complete}  + \Phi
\end{equation}
Eq. (\ref{VIsInKernel}) shows that $v(\lambda,\mu)$ is in the kernel of $Q$.

\paragraph     {Comment on the symmetry of $\Psi$}
Notice that the leading $\omega$-double-trace part $\stackrel{\rm tr}{\Psi}$ is 
{\em antisymmetric} under the exchange of $\mu\leftrightarrow \lambda$, while the leading
$\omega$-traceless part is symmetric under such an exchange.

\paragraph     {Comment on the relation to \cite{Berkovits:2008ga}}
Notice that $\stackrel{\rm tr}{\Psi}$  contains the expression $||\mu \cap \theta\cup\lambda\cap\theta ||$, which is 
$Q_L$-closed and up to a $Q_L$-exact expression equal to  
the dilaton vertex $\mbox{Str}(\lambda_3\lambda_1)$ of \cite{Berkovits:2008ga}. It is multiplied by $\omega\cdots\omega$.
In terms of the ansatz (\ref{Ansatz}) this corresponds to multiplying 
the $\mbox{Str}(\lambda_3\lambda_1)$ by the $(x,\theta)$-dependent profile wave function of the 
excitation (while the dilaton of \cite{Berkovits:2008ga} was constant,
{\it i.e.} $(x,\theta)$-independent). However simply taking:
\begin{equation}
\mbox{Str}(\lambda_3\lambda_1) \mapsto 
\mbox{Str}(\lambda_3\lambda_1) f(x,\theta)
\end{equation}
would not be BRST-closed. This is why we needed to add the $\omega$-traceless part.

\subsection{Could $v(\lambda,\mu)$ be BRST-exact?}
Notice that $v(\lambda,\mu)$ does not have a term quadratic in $\lambda$.
Therefore, the only way it could be $Q$-exact is the following:
\begin{equation}
v(\lambda,\mu) = Q A(\lambda) + Q B(\mu)
\end{equation}
where $A(\lambda)$ and $B(\mu)$ are linear functions of $\lambda$ and $\mu$ respectively, and 
moreover $Q_LA(\lambda) =0$. Then $A(\lambda)$ is $Q_L$-exact: 
\begin{equation}
A(\lambda)=Q_LC
\end{equation}
because we have proven in Section \ref{sec:TwoZeroPart} that the cohomology
of $Q_L$ is trivial. Therefore:
\begin{equation}
v(\lambda,\mu) = Q_RQ_L C + QB(\mu)
\end{equation}
This means that:
\begin{equation}\label{v11}
v(\lambda,\mu)_{1,1} = Q_L(X)
\end{equation}
where $X = -Q_RC + B$. 
Notice that $v(\lambda,\mu)_{1,1}$ is of the type $\lambda\mu\theta\theta$ plus terms of the higher order 
in $\theta$. Since the leading term in $v(\lambda,\mu)_{1,1}$ is nontrivial in $Q_{L+}$-cohomology,
it must be that $X=[\mu\theta] + \ldots$ where the coefficient of $[\mu\theta]$ is nonzero; 
but the expression of the type $[\mu\theta]$ cannot be annihilated by $Q_{L+}$. This 
means that the right hand side of (\ref{v11}) cannot be of they type $[\lambda\mu\theta\theta]$,
because it would necessarily contain the $[\lambda\mu]$ term (term without $\theta$'s).

\paragraph     {Conclusion}
We therefore conclude that the following expression:
\begin{equation}
v (\lambda,\mu) = \Psi_{\rm complete} + \Phi
\end{equation}
is a nontrivial covariant vertex.

\section{Gauge choices}\label{sec:GaugeChoices}
\subsection{Symmetry properties under the exchange $\lambda\leftrightarrow\mu$}
Notice that $Q_+$ is symmetric with respect to the 
exchange $\lambda\leftrightarrow\mu$, and $Q_-$ is antisymmetric. Therefore we have the
following symmetry of the BRST complex:
\begin{equation}\label{ESymmetry}
(Ev)(\lambda,\mu,\theta) = v(\mu,\lambda,i\theta)
\end{equation}
Our vertex, as we constructed it, is a sum of the $\mu\mu$ part and 
the $\lambda\mu$ part; the $\lambda\lambda$ part is zero; this is the ``asymmetric gauge''. 

\subsection{Rocket gauge}
We can do a gauge transformation removing the $\mu\mu$ part, and get the 
``rocket gauge'' where the vertex is purely $\lambda\mu$.
The leading term of such a gauge transformation is:
\begin{equation}
\phi = \leftsup{\bullet}{\{}\mu\stackrel{0}{\cap}\theta\}^{\bullet}\otimes 
(\omega_{\bullet\bullet})^{\otimes (n+1)} \otimes 
(\omega^{\bullet\bullet})^{\otimes n} 
\end{equation}
After this gauge transformation the vertex is:
\begin{align}
\left.v_{\rm rocket}\right|_{\lambda\mu} = & \; 
\leftsup{\bullet}{\{}\mu\stackrel{0}{\cap}\lambda\}^{\bullet}\otimes 
(\omega_{\bullet\bullet})^{\otimes (n+1)} \otimes 
(\omega^{\bullet\bullet})^{\otimes n} + \ldots
\\   
\left.v_{\rm rocket}\right|_{\lambda\lambda} 
= \left.v_{\rm rocket}\right|_{\mu\mu} =  & \; 0
\end{align}
The leading ({\it i.e.} $\theta^0$) term of $\Psi_{\rm rocket}$ is symmetric with respect to the
exchange $\lambda\leftrightarrow\mu$, the $\theta^2$ term is antisymmetric with respect to $\lambda\leftrightarrow\mu$, then 
the $\theta^4$ term is again symmetric and so on. The symmetry w.r.to $\lambda\leftrightarrow\mu$ 
correlates with the power of $\theta$.

\subsection{Airplane gauge}\label{sec:AirplaneGauge}
Consider:
\begin{equation}
v_{\rm air}= {1\over 2}(v+Ev)
\end{equation}
where $Ev$ is defined in (\ref{ESymmetry}). This is equal to:
\begin{align}
v_{\rm air} = & \; {5\over 8}\left({1\over n+1}  \Psi^{\rm tr} + \right.
\nonumber \\   
& \quad + {1\over 2} \;\leftsup{\bullet}{\theta}\cap\lambda\cup\theta\cap
   \theta\cup\lambda\cap\theta^{\bullet}
  \otimes (\omega^{\bullet\bullet})^{\otimes n}
  \otimes (\omega_{\bullet\bullet})^{\otimes (n+1)} +
\nonumber \\   
& \quad + {1\over 2} \;\leftsub{\bullet\;}{\theta}\cup\lambda\cap\theta\cup
   \theta\cap\lambda\cup\theta_{\bullet}
  \otimes (\omega_{\bullet\bullet})^{\otimes n}
  \otimes (\omega^{\bullet\bullet})^{\otimes (n+1)} +
\nonumber \\   
& \quad + {1\over 2} \;\leftsup{\bullet}{\theta}\cap\lambda\cup\theta\cap
   \theta\cup\lambda\cap\theta^{\bullet}
  \otimes (\omega^{\bullet\bullet})^{\otimes n}
  \otimes (\omega_{\bullet\bullet})^{\otimes (n+1)} +
\nonumber \\   
& \quad + {1\over 2} \;\leftsub{\bullet\;}{\theta}\cup\lambda\cap\theta\cup
   \theta\cap\lambda\cup\theta_{\bullet}
  \otimes (\omega_{\bullet\bullet})^{\otimes n}
  \otimes (\omega^{\bullet\bullet})^{\otimes (n+1)} +
\nonumber \\  
& \quad + {1\over 2} \;\leftsup{\bullet}{\theta}\cap\mu\cup\theta\cap
   \theta\cup\mu\cap\theta^{\bullet}
  \otimes (\omega^{\bullet\bullet})^{\otimes n}
  \otimes (\omega_{\bullet\bullet})^{\otimes (n+1)} +
\nonumber \\   
& \quad + {1\over 2} \;\leftsub{\bullet\;}{\theta}\cup\mu\cap\theta\cup
   \theta\cap\mu\cup\theta_{\bullet}
  \otimes (\omega_{\bullet\bullet})^{\otimes n}
  \otimes (\omega^{\bullet\bullet})^{\otimes (n+1)} +
\nonumber \\   
& \quad + {1\over 2} \;\leftsup{\bullet}{\theta}\cap\mu\cup\theta\cap
   \theta\cup\mu\cap\theta^{\bullet}
  \otimes (\omega^{\bullet\bullet})^{\otimes n}
  \otimes (\omega_{\bullet\bullet})^{\otimes (n+1)} +
\nonumber \\   
& 
   \quad + \left. {1\over 2} \;\leftsub{\bullet\;}{\theta}\cup\mu\cap\theta\cup
   \theta\cap\mu\cup\theta_{\bullet}
  \otimes (\omega_{\bullet\bullet})^{\otimes n}
  \otimes (\omega^{\bullet\bullet})^{\otimes (n+1)} +
\right)
\nonumber \\ 
& \quad + \ldots
\label{AirplaneGauge}
\end{align}
where $\ldots$ mean terms of the higher order in $\theta$, more precisely terms of the 
form $\mu\lambda\theta^{\geq 4}$,$\mu\mu\theta^{\geq 6}$ and $\lambda\lambda\theta^{\geq 6}$. Indeed  from (\ref{QOnDoubleTrace0}) we get:
\begin{align}
& Q_{L-}\left( 
||\mu \cap \theta\cup\lambda\cap\theta || \;
(\omega^{\bullet\bullet})^{\otimes (n+1)}\;(\omega_{\bullet\bullet})^{\otimes (n+1)}
\right) =
\nonumber \\[5pt]    
= &  \; - {4\over 5}(n+1) \;
\mu_{[\bullet}\cup\theta\cap\lambda\cup\theta\cap\{\lambda\stackrel{0}{\cup}\theta\}_{\bullet]}\;
(\omega^{\bullet\bullet})^{\otimes (n+1)}\;(\omega_{\bullet\bullet})^{\otimes n} -
\label{QOnDoubleTrace} \\     
& - \; {4\over 5}(n+1) \;
\mu^{[\bullet}\cap\theta\cup\lambda\cap\theta\cup\{\lambda\stackrel{0}{\cap}\theta\}^{\bullet]}
(\omega^{\bullet\bullet})^{\otimes n}\;(\omega_{\bullet\bullet})^{\otimes (n+1)}
\nonumber
\end{align}
On the other hand:
\begin{align}
Q_{R+} & \left(\leftsup{\alpha}{\theta}\cap\lambda\cup\theta\cap
   \theta\cup\lambda\cap\theta^{\beta} 
   - {1\over 4}\; \omega^{\alpha\beta} \;
   ||\theta\cap\lambda\cup\theta\cap\theta\cup\lambda\cap\theta||
\right) \simeq
\nonumber \\    
\simeq &  \; {8\over 5}\mu^{[\alpha}\cap\{\lambda\stackrel{0}{\cup}\theta\}\cap
 \theta\cup\lambda\cap\theta^{\beta]} 
   - {2\over 5}\; \omega^{\alpha\beta} \;
   ||\mu\cap\{\lambda\stackrel{0}{\cup}\theta\}\cap 
   \theta\cup\lambda\cap\theta||
\\   
Q_{R+} & \left(\leftsub{a\;}{\theta}\cup\lambda\cap\theta\cup
   \theta\cap\lambda\cup\theta_{\;b} 
   - {1\over 4}\; \omega_{ab} \;
   ||\theta\cup\lambda\cap\theta\cup\theta\cap\lambda\cup\theta||
\right) \simeq
\nonumber \\    
\simeq &  \; {8\over 5}\mu_{[a}\cup\{\lambda\stackrel{0}{\cap}\theta\}\cup
 \theta\cap\lambda\cup\theta_{b]} 
   - {2\over 5}\; \omega_{ab} \;
   ||\mu\cup\{\lambda\stackrel{0}{\cap}\theta\}\cup 
   \theta\cap\lambda\cup\theta||
\end{align}
\remv{Calculation:}\rem{photo/shot0018.png}
In other words:
\begin{align}
Q_{R+} \left( \theta\cap\lambda\cup\theta\cap
   \theta\cup\lambda\cap\theta \right)^{\bullet\bullet}_{\omega-{\rm less}}
= {8\over 5}\left( \mu\cap\{\lambda\stackrel{0}{\cup}\theta\}\cap
 \theta\cup\lambda\cap\theta
\right)^{[\bullet\bullet]}_{\omega-{\rm less}}
\\[8pt]    
Q_{R+} \left( \theta\cup\lambda\cap\theta\cup
   \theta\cap\lambda\cup\theta \right)_{\bullet\bullet}^{\omega-{\rm less}}
= {8\over 5}\left( \mu\cup\{\lambda\stackrel{0}{\cap}\theta\}\cup
 \theta\cap\lambda\cup\theta
\right)_{[\bullet\bullet]}^{\omega-{\rm less}}
\end{align}
\remv{Check of consistency}\rem{
\begin{align}
& Q_{R+}\left((\theta\cup\lambda\cap\theta)^{[\alpha}_{[a}
(\theta\cup\lambda\cap\theta)^{\beta]}_{b]}\right) =
\nonumber \\    
= & \; - \mu_{[a}\cup\{\lambda\stackrel{0}{\cap}\theta\}^{[\alpha}
\; \theta^{\beta]}\cap\lambda\cup\theta_{b]}
-
\mu^{[\alpha}\cap\{\lambda\stackrel{0}{\cup}\theta\}_{[a}
\; \theta_{b]}\cup\lambda\cap\theta^{\beta]} \simeq
\nonumber \\   
\simeq & \;
{2\over 5} \omega^{\alpha\beta}\mu_{[a}\cup
\{\lambda\stackrel{0}{\cap}\theta\}\cup
\theta\cap\lambda\cup\theta_{b]}
+ {2\over 5}\omega_{ab}\mu^{[\alpha}\cap\{\lambda\stackrel{0}{\cup}\theta\}\cap
 \theta\cup\lambda\cap\theta^{\beta]} 
\label{QROnTLTTLT}
\end{align}
photo/shot0019.png}
Therefore $v_{\rm air}$ is BRST-closed; the BRST variation of the ``fuselage''  $\mu\lambda$  
 is cancelled by the BRST variations of the ``wings'' $\mu\mu$ and $\lambda\lambda$.

\begin{figure}[h!]
\centering
\includegraphics[scale=0.25]{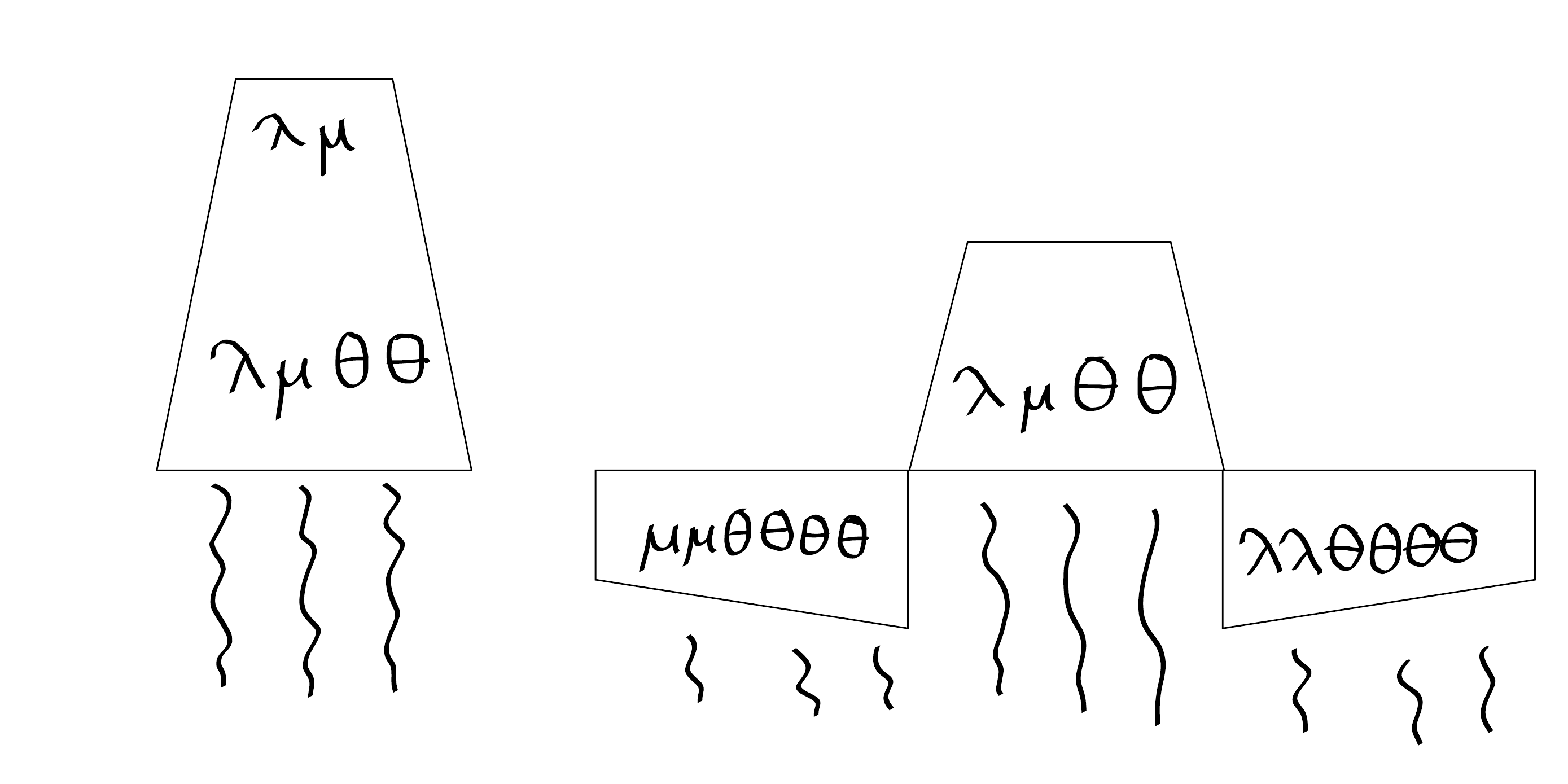}
\caption{The rocket gauge and the airplane gauge; terms of higher order in 
$\theta$ are shown as wavy lines}
\end{figure} 

\section{$B$-field and dilaton for $\Psi = {\bf 1}\otimes V$}\label{sec:BFieldAndDilaton}
\subsection{Simplifications in the special case $\Psi = {\bf 1}\otimes V$}
Using  (\ref{AirplaneGauge}) in (\ref{Ansatz}) with $\Psi = {\bf 1}\otimes V$
and $g$ given by (\ref{G0Gauge}) we get, to the lowest order in $\theta$s:
\begin{equation}
V = \langle v(\lambda,\mu)|_{\theta \to \theta_L + \theta_R} \;,\;
e^x V \rangle
\end{equation}
The effect of substitution $\theta = \theta_L + \theta_R$ is studied in Appendix \ref{sec:ThetaLPlusThetaR}, where
we show that the leading term of the $\theta$-expansion is proportional to:
\begin{equation}\label{OneTimesVLeadingTermOfThetaExpansion}
\langle (\omega^{\bullet\bullet})^{\otimes (n+1)} 
\otimes 
(\omega_{\bullet\bullet})^{\otimes (n+1)} \;,\;e^x V
\rangle \; 
(\lambda\Gamma^m\theta_L)(\mu\Gamma_m\theta_R)
\end{equation}

\subsection{Dilaton profile in the flat space limit}\label{sec:DilatonProfile}
\subsubsection{Spectral sequence of the flat space expansion}
Consider the near flat space expansion with both $x$ 
and $\theta$ scaling like $R^{-1}$. Let us choose the gauge (cp. (\ref{LambdaMinusInTermsOfLambdaPlus}) and (\ref{MuMinusInTermsOfMuPlus})):
\begin{equation}\label{G0Gauge}
g = 
\exp\left(\begin{array}{cc}
      0 & \theta_L + \theta_R \cr
      i(\cap(\theta_L-\theta_R)\cup) & 0
\end{array}\right)
\exp\left(\begin{array}{cc}
   x_A & 0 \cr
   0 & x_S
\end{array}\right)
\end{equation}
In this gauge the expression for  $Q$ up to the order $R^{-1}$ is\footnote{Notice that there is a 
minus sign in front of $(\theta_R\Gamma^m\mu)$. This means that the flat
space limit $\theta_R$ and $\lambda_R$ are actually $i\theta_R$ and $i\mu$.}:
\begin{equation}\label{QApprox}
Q_{\rm approx} = \lambda {\partial\over\partial\theta_L} 
+ \mu {\partial\over\partial\theta_R}
+ \left((\theta_L\Gamma^m\lambda) - (\theta_R\Gamma^m\mu)\right)
{\partial\over\partial x^m}
\end{equation}
The next terms will be of the order $R^{-2}$, for example $\theta\theta\lambda\partial_{\theta}$. Observe that 
with our choice of the gauge (\ref{G0Gauge}) there are no terms of the type $x\lambda{\partial\over\partial\theta}$. 
In other words this approximation of the AdS BRST operator looks in this 
gauge exactly like the flat space BRST operator. 
Moreover, let us consider the following splitting of $Q$:
\begin{equation}
Q = \lambda {\partial\over\partial\theta_L} 
+ \mu {\partial\over\partial\theta_R} + Q_1
\end{equation}
where the first term in the expansion of $Q_1$ is the 
$\left((\theta_L\Gamma^m\lambda) - (\theta_R\Gamma^m\mu)\right)
{\partial\over\partial x^m}$ on the right hand side of (\ref{QApprox}).
Observe that the gauge (\ref{G0Gauge}) leads to a natural grading.
Namely, let $F^p$ denote the space of functions of $(\lambda,\mu,x,\theta_L,\theta_R)$ having
the degree in $\lambda$, plus the degree in $\mu$, plus the degree in
$\theta_L$, plus the degree in $\theta_R$, greater or equal to $2p$. We observe the following
action of operators on grading:
\begin{align}
\lambda {\partial\over\partial\theta_L} 
+ \mu {\partial\over\partial\theta_R} \quad &:\quad
F^p \to F^p 
\\    
Q_1 \quad &: \quad
F^p\to F^{p+1}
\end{align}
This is special to the gauge (\ref{G0Gauge}); in this sense this a good gauge choice 
for the near flat space expansion. Let us calculate the cohomology of  $Q$ 
using the spectral sequence of this filtration. The first page $E_1^{p,q}$ is:
\begin{equation}
E_1^{p,q} = \left.H\left(\lambda {\partial\over\partial\theta_L} 
+ \mu {\partial\over\partial\theta_R}
\; ,\;\; 
F^p\right)\right|_{\mbox{\small degree in $\lambda$ + degree in $\mu$ \;=\; $p+q$}}
\end{equation}
and the first differential $d_1\:\;\: E_1^{p,q} \to E_1^{p+1,q}$ is induced by $Q_1$. 

We see from (\ref{QApprox}) that $d_1$  acts like in flat space. In particular,
this shows that the vertex which we constructed is not $Q$-exact. Indeed,
observe that $E_1^{p,q} = 0$ when $p+q=1$ and $p<1$ (since $q= {1\over 2}(\mbox{gh\#} - \mbox{\#}\theta)$). 
This implies that $E_1^{2,0}$ can only be cancelled by $d_1(E_1^{1,0})$ and not by any 
higher differential $d_{>1}$. But $d_1$ acts as in flat space. We will see now that 
for $n\geq 1$ the flat space limit of our vertex corresponds to a nontrivial 
dilaton profile. This means that it cannot be gauged away in flat space and 
therefore neither in AdS. 

\subsubsection{Polynomial SUGRA solutions in flat space}

To get the flat space limit we expand (\ref{OneTimesVLeadingTermOfThetaExpansion}) in powers of $x$ and keep the 
lowest order terms in $x$. This results in expressions of the type:
\begin{equation}
P(x_A,x_S) \; (\lambda\Gamma^m\theta_L)(\mu\Gamma_m\theta_R)
\end{equation}
where $P(x_A, x_S)$ are harmonic homogeneous polynomials of $x$.

Therefore  in the flat space limit our states become polynomial in 
the coordinates. Notice that the linearized solutions most commonly studied 
in string theory are exponential, of the form $e^{ikx}$. In our opinion, the 
polynomial solutions deserve further investigation. The exponential
solutions factorize into the left and right moving parts, $e^{ikx_L} e^{ikx_R}$.
The polynomial solutions do not factorize. 

\subsection{Dilaton profile in $AdS_5\times S^5$}

We obtain $B_{\mu\nu} = 0$ and $G_{\mu\nu} = \phi(x) \delta_{\mu\nu}$ where:
\begin{equation}
\phi(x) = \omega^{\alpha_1\beta_1} \cdots \omega^{\alpha_{n+1}\beta_{n+1}}\; 
\omega_{a_1b_1} \cdots \omega_{a_{n+1}b_{n+1}}\; 
(g V)_{\alpha_1\beta_1\cdots\alpha_{n+1}\beta_{n+1}}^{a_1b_1\cdots a_{n+1}b_{n+1}}
\end{equation}
In this formula $g=e^x$ parametrizes the bosonic space $AdS_5\times S^5$.

The expression for $\phi$ is more transparent in the vector notations.
Let us think of $V$ as the symmetric traceless tensor of $SO(6)$ (the 
upper latin indices) and the symmetric traceless tensor of $SO(2,4)$
(the lower greek indices).  We parametrize the point of $AdS_5\times S^5$ 
as a pair of vectors  $(X,Y)\in {\bf R}^{2+4}\oplus {\bf R}^6$, $||X||^2 = ||Y||^2 = 1$. We get:
\begin{equation}
\phi(X,Y) = X^{A_1}\cdots X^{A_{n+1}} Y_{I_1} \cdots Y_{I_{n+1}}
V^{I_1\ldots I_{n+1}}_{A_1\ldots A_{n+1}}
\end{equation}

\section{Generalization}\label{sec:Generalization}
The construction of the vertex in the  ``airplane gauge'' allows the following
generalization. 

\subsection{General ansatz}
Suppose that ${\cal H}'$ is such that the second Casimir operator
of ${\bf g}$ vanishes on ${\cal H'}$. Suppose that exists a vector $\Omega\in {\cal H}'$ which is:
\begin{enumerate}
\item annihilated by ${\bf g}_0 = sp(2)_A\oplus sp(2)_S \subset {\bf g}$
\item annihilated by ${\bf n}_-$
\end{enumerate}
The subalgebra ${\bf g}_{\bar{2}}\subset {\bf g}$ is the vector represenation of $sp(2)_A$ 
plus the vector of $sp(2)_S$:
\begin{equation}
{\bf g}_{\bar{2}} = \mbox{Vect}_A + \mbox{Vect}_S
\end{equation}
The generators of ${\bf g}_{\bar{2}}$ will be denoted $t_{[\alpha\beta]}$ and $t^{[ab]}$. Let us consider:
\begin{align}
v(\lambda,\mu) = &\; ||\mu\cap\theta\cup\lambda\cap\theta||\;\Omega \; + 
\nonumber \\   
& \; + \left(
   \theta\cap\lambda\cup\theta\cap\theta\cup\lambda\cap\theta
\right)^{\alpha\beta}_{\omega-{\rm less}}
\;t_{[\alpha\beta]}\Omega \; + 
\nonumber \\
& \; + \left(
   \theta\cup\lambda\cap\theta\cup\theta\cap\lambda\cup\theta
\right)_{ab}^{\omega-{\rm less}}
\;t^{[ab]}\Omega 
\nonumber \\   
& \; + \left(
   \theta\cap\mu\cup\theta\cap\theta\cup\mu\cap\theta
\right)^{\alpha\beta}_{\omega-{\rm less}}
\;t_{[\alpha\beta]}\Omega \; + 
\nonumber \\
& \; + \left(
   \theta\cup\mu\cap\theta\cup\theta\cap\mu\cup\theta
\right)_{ab}^{\omega-{\rm less}}
\;t^{[ab]}\Omega + 
\label{GeneralAirplane}
\\   
&\; + \mbox{ [terms of the  order $\lambda\mu\theta^{\geq 4}$,
             $\lambda^2\theta^{\geq 6}$ and $\mu^2\theta^{\geq 6}$]}
\nonumber
\end{align}
This is the generalization of (\ref{AirplaneGauge}); comparison with (\ref{PsiTrace}) shows that in our
explicit finite-dimensional construction $\Omega$ is the product of the 
Roiban-Siegel symplectic forms:
\begin{equation}\label{OmegaInFiniteCase}
\Omega = \left(\omega^{\bullet\bullet}\right)^{\otimes (n+1)}
\otimes
\left(\omega_{\bullet\bullet}\right)^{\otimes (n+1)}
\end{equation}

\subsection{Deriving the general ansatz in the airplane gauge}
\subsubsection{Constructing the BRST closed expression}
Notice that $Q_+$ is symmetric with respect to the  exchange $\lambda\leftrightarrow\mu$, and $Q_-$ 
is antisymmetric. Therefore we have the $E$-symmetry (\ref{ESymmetry}) of the BRST 
complex. Observe that $E^2 = (-)^{\# \theta}$ and:
\begin{align}
EQ_{L}   = & - i Q_{R} E
\\  
EQ_{R} = & - i Q_{L} E
\\
   EQ_{L} Q_{R} = &\; Q_{L} Q_{R} E
\end{align}
Observe that $||\mu\cap \theta\cup\lambda\cap\theta||$ is meta-odd and $E$-even. We get, as in
Section \ref{sec:AirplaneGauge}:
\begin{align}
& Q_{L} ||\mu\cap \theta\cup\lambda\cap\theta|| \Omega = 
\nonumber \\  
=\; & Q_{R} \left(
   (\theta\lambda\theta\theta\lambda\theta)^{\alpha\beta}T_{\alpha\beta}\Omega 
   + (\theta\lambda\theta\theta\lambda\theta)_{ab}T^{ab}\Omega 
   + Q_LA_{[\lambda\theta^5]} + X_{[\lambda^2\theta^6+\ldots]}
\right) + 
\nonumber \\  
& + \; Y_{[\mu\lambda^2\theta^5+\ldots]}\; 
\label{DefYML2T5}
\end{align}
where $X_{[\lambda^2\theta^6+\ldots]}$ is such that:
\begin{equation}
Q_{L}\left(
   (\theta\lambda\theta\theta\lambda\theta)^{\alpha\beta}T_{\alpha\beta}\Omega 
   + (\theta\lambda\theta\theta\lambda\theta)_{ab}T^{ab}\Omega + X_{[\lambda^2\theta^6+\ldots]}
\right) = 0
\end{equation}
The obstacle to the existence of such  $X_{[\lambda^2\theta^6+\ldots]}$ is:
\begin{equation}
 (\theta\lambda\theta\theta\lambda\theta)^{\alpha\beta}(\theta\lambda)^{\gamma\delta}T_{\alpha\beta}T_{\gamma\delta}\Omega 
   + (\theta\lambda\theta\theta\lambda\theta)_{ab}(\theta\lambda)_{cd}
T^{ab}T^{cd}\Omega
\end{equation}
and therefore it is of the type $\lambda^3\theta^5$. It vanishes when the bosonic quadratic 
Casimir vanishes on $\Omega$; the bosonic quadratic Casimir vanishes on $\Omega$ 
because the full quadratic Casimir of ${\bf g}$ vanishes on ${\cal H}'$, and $\Omega$ is 
annihilated by ${\bf n}_-$.

Notice that $Y_{[\mu\lambda^2\theta^5+\ldots]}$ in Eq. (\ref{DefYML2T5}) is $Q_{L}$-closed and therefore $Q_{L}$-exact:
\begin{equation}\label{YML2T5isQexact}
Y_{[\mu\lambda^2\theta^5+\ldots]} = Q_L  Z_{[\mu\lambda\theta^6+\ldots]}
\end{equation}
Observe that $Q_RQ_L  Z_{[\mu\lambda\theta^6+\ldots]}$ is $E$-even. Therefore we get:
\begin{equation}
Q_L Q_R (Z - EZ) = 0
\end{equation}
This implies the existence of $U_{[\mu^2\theta^6+\ldots]}$ and $V_{[\lambda^2\theta^6+\ldots]}$ such that:
\begin{align}
Q_R(Z - EZ) = & Q_L  U_{[\mu^2\theta^6+\ldots]}
\\  
Q_L (Z - EZ) = & Q_R V_{[\lambda^2\theta^6+\ldots]}
\end{align}
Observe that $Q_L V = 0$ and therefore exists $W_{[\lambda\theta^7+\ldots]}$ such that:
\begin{equation}
V = Q_L  W_{[\lambda\theta^7+\ldots]}
\end{equation}
We get:
\begin{equation}
Q_L (Z - EZ + Q_R W) = 0
\end{equation}
Now Eq. (\ref{YML2T5isQexact}) gives:
\begin{equation}
Y_{[\mu\lambda^2\theta^5+\ldots]} = Q_L  \left(
EZ_{[\mu\lambda\theta^6+\ldots]} - Q_RW_{[\lambda\theta^7+\ldots]}
\right)
\end{equation}
Now we get:
\begin{align}
& Q_L  \left(||\mu\cap \theta\cup\lambda\cap\theta|| \Omega 
- EZ + Q_RW \right)
= 
\nonumber \\  
=\; & Q_R \left(
   (\theta\lambda\theta\theta\lambda\theta)^{\alpha\beta}T_{\alpha\beta}\Omega
   + 
(\theta\lambda\theta\theta\lambda\theta)_{ab}T^{ab}\Omega + Q_LA_{[\lambda\theta^5]} + X_{[\lambda^2\theta^6+\ldots]}
\right) 
\label{QLonModifiedMTLT}
\end{align}
and:
\begin{align}
& Q_R \left(||\mu\cap \theta\cup\lambda\cap\theta|| \Omega 
- EZ + Q_RW \right)
= 
\label{QRonModifiedMTLT}
\\  
=\; & Q_L  \left(
    (\theta\mu\theta\theta\mu\theta)^{\alpha\beta}T_{\alpha\beta}\Omega 
   + (\theta\mu\theta\theta\mu\theta)_{ab}T^{ab}\Omega 
   - i Q_R(EA)_{[\mu\theta^5]} + (EX)_{[\mu^2\theta^6+\ldots]}
\right) 
\nonumber 
\end{align}
The second equality was derived in the following way:
\begin{align}
& Q_R \left(||\mu\cap \theta\cup\lambda\cap\theta|| \Omega 
- EZ + Q_RW \right)
= 
 iE \; Q_L  \left(||\mu\cap \theta\cup\lambda\cap\theta|| \Omega 
- Z \right) =
\nonumber \\  
= & \; iE \; Q_R \left(
   (\theta\lambda\theta\theta\lambda\theta)^{\alpha\beta}T_{\alpha\beta}\Omega 
   + (\theta\lambda\theta\theta\lambda\theta)_{ab}T^{ab}\Omega 
+ Q_LA_{[\lambda\theta^5]}
+ X_{[\lambda^2\theta^6+\ldots]}
\right) = 
\nonumber \\  
= & \; Q_L  \left(
   (\theta\mu\theta\theta\mu\theta)^{\alpha\beta}T_{\alpha\beta}\Omega 
   + (\theta\mu\theta\theta\mu\theta)_{ab}T^{ab}\Omega 
- i Q_R(EA)_{[\mu\theta^5]}
+ (EX)_{[\mu^2\theta^6+\ldots]}
\right) 
\nonumber
\end{align}
Let us therefore consider the following expression:
\begin{align}
& ||\mu\cap \theta\cup\lambda\cap\theta|| \Omega 
- EZ + Q_RW -
\nonumber
\\ 
- & \left(
   (\theta\lambda\theta\theta\lambda\theta)^{\alpha\beta}T_{\alpha\beta}\Omega 
   + (\theta\lambda\theta\theta\lambda\theta)_{ab}T^{ab}\Omega + X_{[\lambda^2\theta^6+\ldots]} + Q_LA_{[\lambda\theta^5]}
\right) -
\\  
- & \left(
    (\theta\mu\theta\theta\mu\theta)^{\alpha\beta}T_{\alpha\beta}\Omega 
   + (\theta\mu\theta\theta\mu\theta)_{ab}T^{ab}\Omega 
   + (EX)_{[\mu^2\theta^6+\ldots]} - i Q_R(EA)_{[\mu\theta^5]}
\right) 
\nonumber
\end{align}
Eqs. (\ref{QLonModifiedMTLT}) and (\ref{QRonModifiedMTLT}) imply that this expression is $Q$-closed. 

Also observe that the
terms $Q_LA_{[\lambda\theta^5]}$ and $i Q_R(EA)_{[\mu\theta^5]}$ are gauge equivalent to expressions
 of the form $[\lambda\mu\theta^4 + \ldots]$.

\subsubsection{Showing that the constructed expression is not BRST exact}
Let us prove that it is not $Q$-exact. Assume that we have found $\phi$ such 
that $Q\phi = v$. Then in particular:
\begin{align}
Q_L\phi_{\lambda} = 
& \;  \left(
   \theta\cap\lambda\cup\theta\cap\theta\cup\lambda\cap\theta
\right)^{\alpha\beta}_{\omega-{\rm less}}
\;t_{[\alpha\beta]}\Omega \; + 
\nonumber \\
& \; + \left(
   \theta\cup\lambda\cap\theta\cup\theta\cap\lambda\cup\theta
\right)_{ab}^{\omega-{\rm less}}
\;t^{[ab]}\Omega  + \ldots
\end{align}
Such $\phi_{\lambda}$ should start with the leading term $\lambda\theta$. But $Q_{R+}$ on the leading 
term will then give $\lambda\mu$. This means that $\phi$ in fact does not gauge away $v$, 
but actually rather brings it  to the ``rocket'' gauge ({\it i.e.} removes
the $\lambda\lambda$ and $\mu\mu$-parts at the expense of introducing
the $\theta$-less term in the $\lambda\mu$-part).

Another proof can be given by observing the nontrivial dilaton profile, 
using the methods of Section \ref{sec:DilatonProfile}.

\subsection{Supergravity meaning of $\Omega$}
This generalization of our construction described in Section \ref{sec:Generalization} also works 
for infinite-dimensional representations. But for infinite-dimensional 
representations the construction of $\Omega$ is less transparent than the explicit 
formula (\ref{OmegaInFiniteCase}).

There is a candidate $\Omega$ for ${\cal H}$ being the space of all linearized 
supergravity solutions. Let us think of ${\cal H}'$ as the space of all 
gauge-invariant SUGRA operators at the fixed point of $AdS_5\times S^5$. 
We can restrict ourselves to evaluating them on a particular 
subspace. The supersymmetry transformations of the supergravity fields
can be found {\it e.g.} in \cite{Green:1987mn}. In particular, the transformation laws
for the dilaton-axion field $V_-^{\alpha}$ is:
\begin{equation}\label{GSW}
\delta V_-^{\alpha} = \kappa V_+^{\alpha} \overline{\eta} \lambda^*
\end{equation}
Take a gauge invariant combination, for example $V^1_-/V^2_-$. Then  (\ref{GSW}) implies 
that this combination is invariant under the (complexified) supersymmetry 
transformations which have $\overline{\eta} = 0$ (and the only nonzero parameter is $\overline{\eta}^*$).

This means that this operator is annihilated by ${\bf n}_-$. Therefore we can take
$\Omega$ in the following form:
\begin{align}
\Omega\; & : \; {\cal H}\to {\bf C} 
\nonumber \\   
\Omega\left(
   \begin{array}{c} 
      \mbox{\small \tt SUGRA }\cr 
      \mbox{\small \tt solution} 
   \end{array}
\right)\; & = \;  \left[
   \begin{array}{l} 
      \mbox{\small \tt fluctuation of }V_-^1/V_-^2 \cr
      \mbox{\small \tt evaluated on }
      \mbox{\small \tt this solution} \cr
      \mbox{\small \tt at the marked point } 
      \mbox{\small \tt of }AdS_5\times S^5
   \end{array}
\right]
\end{align}
Here ``fluctuation'' of the field means the difference with the value in the
undeformed $AdS_5\times S^5$. This is, essentially, a complex linear combination of 
the fluctuation of dilaton plus axion:
\begin{equation}
\Omega = \delta \phi + i\psi
\end{equation}
Then we can construct the vertex using (\ref{GeneralAirplane}).

\section{Subspaces and factorspaces}\label{sec:SubspacesAndFactorspaces}
\subsection{Equivalence relation?}\label{sec:EquivalenceRelation}
As we explained in Section \ref{sec:RelatedT} our space ${\cal H}$ is not irreducible.
The kernel of the map $\mbox{ev}:\; {\cal H}\to {\cal T}$ is an invariant subspace, and there is 
no complementary subspace. Because the space of deformations is not
a unitary representation, there is no obvious reason why it should be 
irreducible. It is natural to ask the following question:
\begin{itemize}
\item Is it possible to define the vertex on the irreducible representation
 ${\cal T} = {\cal H}/\mbox{ker(ev)}$ (the Young diagramm representation) rather than ${\cal H}$?
\end{itemize}
This would be possible iff the covariant vertex $v(\lambda,\mu)$ which we constructed 
were $Q$-exact up to $\left(\mbox{ker}(\mbox{ev})\right)^{\perp}$:
\begin{equation}\label{EquivalenceRelation}
v \; \stackrel{?}{\in} \;\mbox{Im}(Q) \; + \; \left(\mbox{ker}(\mbox{ev})\right)^{\perp}
\end{equation}
--- see Eq. (\ref{DefKerEvPerp}). If this conjecture was true, then this would imply that 
the ``dressed'' vertex $V[\Psi](g,\lambda)$ given by (\ref{Ansatz}) is $Q$-exact when $\Psi$ is 
in $\mbox{ker}(\mbox{ev})$. This would imply that the space of linearized deformations is 
really ${\cal T}$ of (\ref{DefT}) rather than ${\cal H}$ of (\ref{DefH}).

However the hypothesis (\ref{EquivalenceRelation}) is not true. Indeed, let us consider 
the ``airplane'' gauge of Eq. (\ref{AirplaneGauge}). We can remove the $\lambda\lambda$ and $\mu\mu$ parts 
(``the wings'') mod $\left(\mbox{ker}(\mbox{ev})\right)^{\perp}$. Indeed, let us look at the structure of  
the $\lambda\lambda$ wing. It consists of the terms like this one:
\begin{equation}\label{SomeThetasWithoutFreeIndices}
\;\leftsup{\bullet}{\theta}\cap\lambda\cup\theta\cap
   \theta\cup\lambda\cap\theta^{\bullet}
  \otimes (\omega^{\bullet\bullet})^{\otimes n}
  \otimes (\omega_{\bullet\bullet})^{\otimes (n+1)} 
\end{equation}
This is $Q_{L+}$-equivalent to 
\begin{equation}\label{AllThetasHaveFreeIndex}
\;\leftsup{\bullet}{\theta}\cap\lambda\cup\theta_{\bullet}
  \;\leftsub{\bullet\;} \theta\cup\lambda\cap\theta^{\bullet}
  \otimes (\omega^{\bullet\bullet})^{\otimes (n+1)}
  \otimes (\omega_{\bullet\bullet})^{\otimes (n+1)} 
\end{equation}
Here all the $\theta$'s enter with one uncontracted index, therefore
this is in $\left(\mbox{ker}(\mbox{ev})\right)^{\perp}$. Therefore:
\begin{equation}
v =  {5\over 8n+8} ||\mu \cap \theta\cup\lambda\cap\theta || 
\;
\left(\omega^{\bullet\bullet}\right)^{\otimes (n+1)}
\otimes
\left(\omega_{\bullet\bullet}\right)^{\otimes (n+1)}  
 \;\mbox{mod }\mbox{(ker(ev))}^{\perp}
\end{equation}
Notice that this expression is BRST closed modulo $\mbox{(ker(ev))}^{\perp}$, but not BRST 
exact modulo $\mbox{(ker(ev))}^{\perp}$. Therefore the hypothesis (\ref{EquivalenceRelation}) is false. 

\paragraph     {Proof that $v$ is not BRST exact modulo $\mbox{(ker(ev))}^{\perp}$}
By the symmetries the only candidate for $Q^{-1}v$ would be:
\begin{equation}\label{ProposedQInverse}
|| \mu\cap\theta || 
\;
\left(\omega^{\bullet\bullet}\right)^{\otimes (n+1)}
\otimes
\left(\omega_{\bullet\bullet}\right)^{\otimes (n+1)}  
-
|| \lambda\cap\theta || 
\;
\left(\omega^{\bullet\bullet}\right)^{\otimes (n+1)}
\otimes
\left(\omega_{\bullet\bullet}\right)^{\otimes (n+1)}  
\end{equation}
(All expressions in this paragraph are mod $\mbox{(ker(ev))}^{\perp}$.)
Consider the $\lambda\lambda$ part of $Q$ of (\ref{ProposedQInverse}):
\begin{align}\label{LambdaLambdaPartModulo}
||\lambda\cap\theta|| \left( 
\{\lambda\cap\theta\}^{\bullet\bullet}\omega_{\bullet\bullet} -
\{\lambda\cup\theta\}_{\bullet\bullet}\omega^{\bullet\bullet} \right)
\otimes (\omega^{\bullet\bullet}\otimes\omega_{\bullet\bullet})^{\otimes n}
\end{align}
But this is not $Q_{L+}$-exact. Indeed, by symmetries, the only candidates are
\begin{align}
||\lambda\cap\theta\cup\theta\cap\theta||\;
(\omega^{\bullet\bullet}\otimes\omega_{\bullet\bullet})^{\otimes (n+1)}
\nonumber\\   
\left(\theta^{\bullet}\cap \lambda\cup\theta \cap\theta^{\bullet} -
\theta^{\bullet}\cap \theta\cup\lambda \cap\theta^{\bullet}\right)
\left(\omega^{\bullet\bullet}\right)^{\otimes n}
\otimes
\left(\omega_{\bullet\bullet}\right)^{\otimes (n+1)}
\nonumber\\
\left(\lambda^{\bullet}\cap \theta\cup\theta \cap\theta^{\bullet} -
\theta^{\bullet}\cap \theta\cup\theta \cap\lambda^{\bullet}\right)
\left(\omega^{\bullet\bullet}\right)^{\otimes n}
\otimes
\left(\omega_{\bullet\bullet}\right)^{\otimes (n+1)}
\nonumber\\   
\left(\theta_{\bullet}\cup \lambda\cap\theta \cup\theta_{\bullet} -
\theta_{\bullet}\cup \theta\cap\lambda \cup\theta_{\bullet}\right)
\left(\omega^{\bullet\bullet}\right)^{\otimes (n+1)}
\otimes
\left(\omega_{\bullet\bullet}\right)^{\otimes n}
\nonumber\\
\left(\lambda_{\bullet}\cup \theta\cap\theta \cup\theta_{\bullet} -
\theta_{\bullet}\cup \theta\cap\theta \cup\lambda_{\bullet}\right)
\left(\omega^{\bullet\bullet}\right)^{\otimes (n+1)}
\otimes
\left(\omega_{\bullet\bullet}\right)^{\otimes n}
\end{align}
but they do not give the right expression when acted on by $Q_{L+}$ (in fact most
of them are $Q_{L+}$-exact).

This proof does not work for the generalized construction described in 
Section \ref{sec:Generalization}, because we do not know how to generalize the step leading
from Eq. (\ref{SomeThetasWithoutFreeIndices}) to Eq. (\ref{AllThetasHaveFreeIndex}). 

\subsection{Vertex for irreducible representations}\label{sec:VertexForIrreps}
We get the short exact sequence:
\begin{equation}
0 \rightarrow \mbox{ker(ev)} \rightarrow {\cal H} \rightarrow {\cal T} 
\rightarrow 0
\end{equation}
Restricting our vertex on $\mbox{ker(ev)}$ we get a nontrivial element of
$H^2(Q,\mbox{ker(ev)})$. If $\mbox{ker(ev)}$ contains an invariant subspace, then
we can repeat the process: 

\paragraph     {The reduction process.}
Generally speaking, suppose that we have an exact sequence of representations:
\begin{equation}
0 \rightarrow A \rightarrow B \rightarrow C \rightarrow 0
\end{equation}
Suppose that we have constructed a universal covariant vertex transforming
in $B$.  Then there are 2 possibilities:
\begin{enumerate}
\item everything in $A$ is exact (an ``equivalence relation'' like in 
   Section \ref{sec:EquivalenceRelation}), or
\item there is no such equivalence relation, the restriction of the
   universal vertex to the states in $A$ is nontrivial in cohomology
\end{enumerate}
In the first case we get the vertex in $C =B/A$, and in the second case we 
get the vertex in $A$. 

Eventually, repeating the process, we obtain a universal vertex for some
irreducible representation. The realization of this process requires the 
detailed study of the structure of the Kac module along the lines of \cite{Gotz:2005jz,Gotz:2006qp,Schomerus:2005bf,Saleur:2006tf,Troost:2011fd} 
and references therein.

\section{Open questions}\label{sec:OpenQuestions}
\begin{enumerate}
\item Understand the field theory side.
\item We have constructed the unintegrated vertex. It would be interesting 
   to carry out the descent procedure and construct the integrated vertex,
   as was done in \cite{Bedoya:2010qz} for the $\beta$-deformation vertex.
\item Measure the supergravity fields corresponding to $v(\lambda,\mu)$ for the 
   states $\Psi$ more general than those studied in Section \ref{sec:BFieldAndDilaton}.
\item The generalization to infinite-dimensional representations described
   in Section \ref{sec:Generalization} requires further study.
\item When $n=0$, what is the relation between the vertex constructed in this
   paper and the vertex of \cite{Bedoya:2010qz}? (See Appendix \ref{sec:BetaDef}.)
\item Generally speaking, it would be interesting to study the vertex 
   operators even in {\em flat space} (in pure spinor, 
   or Green-Schwarz, or NSR formalism), corresponding to the linearized
   SUGRA solutions {\em polynomial} in the coordinates. This would be
   the flat space limit of our construction, as discussed in Section \ref{sec:DilatonProfile}.
\end{enumerate}

\section*{Acknowledgments}
We would like to thank N.~Berkovits and V.~Pershin for useful discussions.
This work was supported in part by the Ministry of Education and Science of 
the Russian Federation under contract 14.740.11.0347, and in part by the RFFI 
grant 10-02-01315.
\appendix
\section{Gamma-matrix expressions}
\label{sec:GammaMatrixExpressions}
\subsection{Correcting $\stackrel{0}{\Psi}_{\urmbf{as}}$}
Consider:
\begin{equation}
\stackrel{0}{\Psi}_{\urmbf{as}} = \left(\mu \Gamma_{\urmbf{a}}\Gamma_{\urmbf{s}} \Gamma_m \theta\right) 
\left( \lambda\Gamma^m \theta \right)
\end{equation}
\begin{equation}
\stackrel{0}{\Xi}_{\;\urmbf{as}} = {1\over 2}(\theta\Gamma_{\urmbf{a}}\Gamma_{\urmbf{s}}\Gamma_m\theta)(\mu\Gamma^m\theta)
\end{equation}
Then we have:
\begin{align}
\stackrel{0}{\Psi}_{\;\urmbf{as}} + Q_{L+}\stackrel{0}{\Xi}_{\;\urmbf{as}}= &\; 
(\mu\Gamma_{\urmbf{a}}\Gamma_{\urmbf{s}}\Gamma^m\theta)(\lambda\Gamma_m\theta) + 
{1\over 2} (\lambda\Gamma_{\urmbf{a}}\Gamma_{\urmbf{s}}\Gamma^m\theta)(\mu\Gamma_m\theta) -
\nonumber \\   
& \; 
- {1\over 2}(\theta\Gamma_{\urmbf{a}}\Gamma_{\urmbf{s}}\Gamma^m\lambda)(\mu\Gamma_m\theta)
+ {1\over 2}(\theta\Gamma_{\urmbf{a}}\Gamma_{\urmbf{s}}\Gamma^m\theta)(\mu\Gamma_m\lambda)
\end{align}
Notice that:
\begin{align}
{1\over 2}(\lambda\Gamma_{\urmbf{a}}\Gamma_{\urmbf{s}}\Gamma^m\theta)(\mu\Gamma_m\theta) -
{1\over 2}(\theta\Gamma_{\urmbf{a}}\Gamma_{\urmbf{s}}\Gamma^m\lambda)(\mu\Gamma_m\theta) =
\nonumber \\    
= (\lambda\Gamma_{\urmbf{a}}\Gamma_{\urmbf{s}}\Gamma^m\theta)(\mu\Gamma_m\theta) -
(\theta\Gamma_{\urmbf{a}}\lambda)(\mu\Gamma_{\urmbf{s}}\theta) -
(\theta\Gamma_{\urmbf{a}}\mu)(\lambda\Gamma_{\urmbf{s}}\theta)
\end{align}
\rem{ photo/shot0008.png } 
Therefore:
\begin{align}
\stackrel{0}{\Psi}_{\urmbf{as}} + Q_L\stackrel{0}{\Xi}_{\;\urmbf{as}}= &\; 
(\mu\Gamma_{\urmbf{a}}\Gamma_{\urmbf{s}}\Gamma^m\theta)(\lambda\Gamma_m\theta) 
- (\theta\Gamma_{\urmbf{a}}\lambda)(\mu\Gamma_{\urmbf{s}}\theta) 
+ {1\over 4} (\theta\Gamma_{\urmbf{a}}\Gamma_{\urmbf{s}}\Gamma^m\theta)(\mu\Gamma_m\lambda) +
\nonumber \\   
& \; + (\lambda\leftrightarrow\mu)
\end{align}
We conclude that $\stackrel{0}{\Psi}_{\urmbf{as}} + Q_L\stackrel{0}{\Xi}_{\;\urmbf{as}}$ is symmetric with respect to $(\lambda\leftrightarrow\mu)$. 
We can also rewrite it as follows:
\begin{align}
\stackrel{0}{\Psi}_{\urmbf{as}} + Q_L\stackrel{0}{\Xi}_{\;\urmbf{as}} = &\; 
6(\theta\Gamma_{[\;\urmbf{a}}\lambda)(\mu\Gamma_{\urmbf{s}]}\theta) +
{3\over 2} (\theta\Gamma_{\urmbf{a}}\Gamma_{\urmbf{s}}\Gamma_m\theta)(\lambda\Gamma^m\mu) 
\end{align}
\rem{ photo/shot0010.png }
\rem{ photo/shot0011.png }
\rem{ photo/shot0012.png }

\subsection{With $\theta=\theta_L + \theta_R$}
\label{sec:ThetaLPlusThetaR}
Now let us investigate the trace part:
\begin{equation}
(\lambda\Gamma^m\theta)(\theta\Gamma_m\mu)
\end{equation}
We substitute $\theta = \theta_L + \theta_R$. Notice that the $\theta_L\theta_L$ part is $Q$-exact:
\begin{equation}
(\lambda\Gamma^m\theta_L)(\theta_L\Gamma_m\mu) = 
Q\left((\lambda\Gamma^m\theta_L)(\theta_L\Gamma_m\theta_R)\right)
\end{equation}
and similarly is the $\theta_R\theta_R$ term. The $\theta_R\theta_L$ part is:
\begin{align}
\; (\lambda\Gamma^m\theta_L)(\theta_R\Gamma_m\mu) +
(\lambda\Gamma^m\theta_R)(\theta_L\Gamma_m\mu) 
\end{align}
Observe:
\begin{align}
& (\lambda\Gamma^m\theta_R)(\theta_L\Gamma_m\mu) =
 \; Q\left( 
   (\lambda\Gamma^m\theta_R)(\theta_L\Gamma_m\theta_R) 
\right) - (\lambda\Gamma^m\mu)(\theta_L\Gamma_m\theta_R) =
\nonumber \\   
= & \; Q\left( 
   (\lambda\Gamma^m\theta_R)(\theta_L\Gamma_m\theta_R) 
\right) + (\mu\Gamma^m\theta_L)(\lambda\Gamma_m\theta_R)
+ (\lambda\Gamma^m\theta_L)(\mu\Gamma_m\theta_R)
\end{align}
which implies that:
\begin{equation}
(\lambda\Gamma^m\theta_R)(\theta_L\Gamma_m\mu) =
{1\over 2}(\lambda\Gamma^m\theta_L)(\mu\Gamma_m\theta_R) + Q(\ldots)
\end{equation}
Therefore:
\begin{equation}
(\lambda\Gamma^m\theta)(\theta\Gamma_m\mu) = 
{3\over 2}(\lambda\Gamma^m\theta_L)(\mu\Gamma_m\theta_R) + Q(\ldots)
\end{equation}

\section{Beta deformation}\label{sec:BetaDef}
Here we will rewrite the $\beta$-deformation vertex of \cite{Bedoya:2010qz} using our current
notations.

We start with the $\lambda\lambda$ part:
\begin{equation}
(V_{LL})^{\alpha\beta}_{ab} = \; \theta^{[\alpha}\cap\lambda\cup\theta_{[a} 
\; \theta_{b]} \cup\lambda\cap\theta^{\beta]}
\end{equation}
Note that this is the complete expression, annihilated by $Q_L$, there is
no need to add the terms of the higher order in $\theta$. 

Now let us proceed to the $\lambda\mu$ part. We act on $V_{LL}$ by $Q_R$ and see if it is 
cancelled by $Q_L$ of something. Notice that $Q_{R-}V_{LL}=0$, and therefore
it is enough to calculate $Q_{R+}V_{LL}$:
\begin{align}
(Q_{R}V_{LL})^{\alpha\beta}_{ab} = (Q_{R+}V_{LL})^{\alpha\beta}_{ab} = & \phantom{-}\; 2\; \mu^{[\alpha}\cap\lambda\cup\theta_{[a} 
\; \theta_{b]} \cup\lambda\cap\theta^{\beta]} - 
\nonumber \\    &
-\; 2\; \theta^{[\alpha}\cap\lambda\cup\mu_{[a} 
\; \theta_{b]} \cup\lambda\cap\theta^{\beta]}
\end{align}
We observe:
\begin{align}
(Q_{R}V_{LL})^{\alpha\beta}_{ab} = & \;
\; Q_{L} \left(2\;\mu^{[\alpha}_{[a} \; \; \theta_{b]} \cup\lambda\cap\theta^{\beta]} \right)
\end{align}
On the other hand:
\begin{align}
Q_{R}\left(2\mu^{[\alpha}_{[a} \; \; \theta_{b]} \cup\lambda\cap\theta^{\beta]} \right) = & 
\phantom{-}\; 2\;\mu^{[\alpha}_{[a} \;\; \mu_{b]}\cup\lambda\cap\theta^{\beta]} - 
\nonumber \\    &
-\;2\;\mu^{[\alpha}_{[a} \; \; \theta_{b]} \cup\lambda\cap\mu^{\beta]} 
\end{align}
Finally: 
\begin{equation}
\; 2 \mu^{[\alpha}_{[a} \;\; \mu_{b]}\cup\lambda\cap\theta^{\beta]} 
-\;2\;\mu^{[\alpha}_{[a} \; \; \theta_{b]} \cup\lambda\cap\mu^{\beta]} 
= \; Q_L\; \left(\mu^{[\alpha}_{[a} \;\; \mu^{\beta]}_{b]}\right)
\end{equation}
Therefore the following expression is $Q$-closed:
\begin{align}\label{VertexForBeta}
\mu^{[\alpha}_{[a} \;\; \mu^{\beta]}_{b]} 
- 2\mu^{[\alpha}_{[a} \; \; \theta_{b]} \cup\lambda\cap\theta^{\beta]}
+ \theta^{[\alpha}\cap\lambda\cup\theta_{[a} 
\; \theta_{b]} \cup\lambda\cap\theta^{\beta]}
\end{align}
There is also a symmetric version:
\begin{align}\label{VertexForBetaSymmetric}
& \mu^{[\alpha}_{[a} \;\; \mu^{\beta]}_{b]} 
+ \lambda^{[\alpha}_{[a} \;\; \lambda^{\beta]}_{b]} 
+ 2\lambda^{[\alpha}_{[a} \; \; \theta_{b]} \cup\mu\cap\theta^{\beta]}
- 2\mu^{[\alpha}_{[a} \; \; \theta_{b]} \cup\lambda\cap\theta^{\beta]} +
\nonumber \\    
+\;&  \theta^{[\alpha}\cap\lambda\cup\theta_{[a} 
\; \theta_{b]} \cup\lambda\cap\theta^{\beta]}
+ \theta^{[\alpha}\cap\mu\cup\theta_{[a} 
\; \theta_{b]} \cup\mu\cap\theta^{\beta]}
\end{align}
This expression is BRST-equivalent to the vertex operator of the 
$\beta$-deformation studied in \cite{Bedoya:2010qz}. Indeed, the vertex operator of \cite{Bedoya:2010qz} is:
\begin{align}
V_{\rm beta} = & \; (\lambda_3 - \lambda_1)\wedge (\lambda_3 - \lambda_1) =
\nonumber \\[5pt]   
= & \; (\lambda_{3+} - \lambda_{1+} + \lambda_{3-} - \lambda_{1-})
\wedge (\lambda_{3+} - \lambda_{1+} + \lambda_{3-} - \lambda_{1-}) 
\end{align}
On the other hand, notice that the following expressions are both $Q$-exact: 
\begin{align}
X = & \; (\lambda_{1+} - \lambda_{1-} + \lambda_{3+} - \lambda_{3-})
\wedge   (\lambda_{1+} + \lambda_{1-} - \lambda_{3+} - \lambda_{3-})
\end{align}
and
\begin{align}
Y = & \; (\lambda_{1+} - \lambda_{1-} + \lambda_{3+} - \lambda_{3-})
\wedge (\lambda_{1+} - \lambda_{1-} + \lambda_{3+} - \lambda_{3-})
\end{align}
We observe that $V_{\rm beta} + 2X + Y$ equals 4 times (\ref{VertexForBeta}). 
\remv{Calculation:}\rem{ photo/100_0633.JPG }
This shows that
indeed (\ref{VertexForBeta}) is identified with the vertex of \cite{Bedoya:2010qz}. However (\ref{VertexForBeta}) does not 
coincide with the particular case of our general construction (\ref{GeneralAirplane}) 
specified to $n=0$. It appears that for $n=0$ we have two different 
vertices, namely (\ref{GeneralAirplane}) and (\ref{VertexForBeta}). The relation between the two remains
to be investigated. We suspect that they both correspond to the $\beta$-deformation,
but in different gauges.

\def\cprime{$'$} \def\cprime{$'$}
\providecommand{\href}[2]{#2}\begingroup\raggedright\endgroup

% \bibliographystyle{jhep} \renewcommand{\refname}{Bibliography}
% \addcontentsline{toc}{section}{Bibliography}
% \bibliography{../andrei}

\end{document}